\documentclass[12pt,dvips]{article}

\usepackage{rotating}
\usepackage{axodraw} 
\usepackage{epsfig}  
\usepackage{epsf}    
\usepackage{psfig}   
\usepackage{cite}    

\def\theequation{\arabic{section}.\arabic{equation}}

\newcommand{\imag}{\Im {\rm m}}
\newcommand{\real}{\Re {\rm e}}

\def\lsim{\:\raisebox{-0.5ex}{$\stackrel{\textstyle<}{\sim}$}\:}

\begin{document}

\renewcommand{\thefootnote}{\fnsymbol{footnote}}

\begin{flushright}
CERN-PH-TH/2004-232 \\
MC-TH-2004-04 \\
{\tt hep-ph/0411379} \\
November 2004
\end{flushright}

\begin{center}
{\bf {\Large Resonant CP Violation in MSSM Higgs Production}}\\[3mm]
{\bf {\Large and Decay at $\gamma \gamma$ Colliders}}
\end{center}

\bigskip\bigskip

\begin{center}
{\large John Ellis$^{\,a}$, Jae Sik Lee$^{\,b}$
                                       and Apostolos Pilaftsis$^{\,b}$}
\end{center}

\begin{center}
{\em $^a$Theory Division, Physics Department, CERN, CH-1211 Geneva 23,
Switzerland}\\[2mm]
{\em $^b$School of Physics and Astronomy, University of Manchester}\\
{\em Manchester M13 9PL, United Kingdom}
\end{center}

\bigskip\bigskip\bigskip

\centerline{\bf ABSTRACT}

\noindent
We study CP-violating phenomena in the production, mixing and decay of
a  coupled  system of CP-violating    neutral Higgs bosons  at $\gamma
\gamma$  colliders, assuming a   Minimal Supersymmetric Standard Model
(MSSM)  Higgs sector in which CP  violation  is radiatively induced by
phases in  the  soft     supersymmetry-breaking  gaugino  masses   and
third-generation trilinear squark couplings. We discuss CP asymmetries
in  the production  and decays of  $\mu^+\mu^-$,
$\tau^+  \tau^-$, ${\bar b} b$  and
${\bar t} t$ pairs.  We find large asymmetries when two (or all three)
neutral Higgs   bosons are  nearly   degenerate with  mass differences
comparable to   their  decay  widths, as   happens naturally   in  the
CP-violating MSSM  for values  of  $\tan  \beta  \stackrel{>}{{}_\sim}
5~(30)$ and large (small) charged Higgs-boson masses.

\newpage

\setcounter{equation}{0}
\section{Introduction}
\label{sec:introduction}

The   Minimal    Supersymmetric  extension  of   the  Standard   Model
(MSSM)~\cite{HPN} offers many  possible sources of CP violation beyond
the  single   Kobayashi--Maskawa    phase  in   the    Standard  Model
(SM). However,  if the   soft supersymmetry-breaking parameters  $m_0,
m_{1/2}$  and $A$ are universal, only  two new  physical CP-odd phases
remain:  one  in the trilinear  couplings $A$  and one in  the gaugino
masses  $m_{1/2}$.  If  Nature  is  described  by  such  a constrained
CP-violating version   of the   MSSM,  the production  and  decays  of
sparticles  would offer    many   direct  signatures  of  these    new
CP-violating          parameters         at                high-energy
colliders~\cite{CPdirect,CPsoft1,CPsoft2}.    Additional  indirect   signatures
could  be   provided by    their  radiative  effects   on    the Higgs
sector~\cite{APLB},       electric          dipole             moments
(EDMs)~\cite{EDM1,EDM2,CKP}              and                 $B$-meson
observables~\cite{Bmeson1,DP}.

The Higgs sector of the MSSM  is affected, to a greater extent, by the
trilinear           phase           at          the           one-loop
level~\cite{APLB,PW,Demir,CDL,CEPW,INhiggs,KW,HeinCP,CEPW2,Maria},
and, to  a lesser extent, by the  gaugino mass phases at  the one- and
two-loop  levels~\cite{CDL,CEPW,INhiggs,CEPW2}.  A  complete treatment
of this loop-induced CP  violation involves a careful consideration of
the three-way mixing between the  CP-even Higgses $h,H$ and the CP-odd
Higgs  boson $A$,  including  off-diagonal absorptive  effects in  the
resummed Higgs-boson propagator matrix~\cite{APNPB,ELP1}. Many studies
have been made of the  masses, couplings, production and decays of the
resulting mixed-CP  Higgs bosons $H_{1,2,3}$, with a  view to searches
at    LEP~\cite{CPX}    and   future    colliders,    such   as    the
LHC~\cite{CHL,CPX,CEMPW,CPpp,CFLMP,KMR},  the  International  $e^+e^-$
Linear      Collider      (ILC)~\cite{CPee},     a      $\mu^+\mu^-$
collider~\cite{CPmumu}  and a $\gamma 
\gamma$ collider~\cite{CPphoton,PPTT,GKS,CKLZ}.
The  main  purpose  of  this  paper is  to  extend the  treatment  of   
three-way  mixing   given  previously in~\cite{ELP1} to  $\gamma\gamma$
colliders.   

The $\gamma\gamma$ colliders offer
unique capabilities for probing  CP violation in the MSSM
Higgs sector, because one  may vary the initial-state polarizations as
well as measure  the   polarizations of  some  final  states  in Higgs
decays.   We    illustrate  these   capabilities    by     considering
coupled-channel  $H_{1,2,3}$ mixing in $\mu^+ \mu^-$, $\tau^+   \tau^-$, 
${\bar b} b$
and ${\bar t} t$  final  states.  Even  with quite small  CP-violating
phases, sizable CP-violating  effects are possible when $\tan\beta$ is
large and/or the charged Higgs  boson mass is large,   so that two  or
three Higgs bosons are nearly degenerate, as we demonstrate in a couple
of specific scenarios.   One of  these  exhibits large  mixing between
three near-degenerate MSSM Higgs  bosons  $H_{1,2,3}$, and the   other
scenario  features  one lighter Higgs   boson  $H_{1}$ and two heavier
states $H_{2,3}$.

The layout  of  this paper  is as follows.   Section~2 provides  basic
formulae for  the   $\gamma  \gamma \rightarrow    {\bar f} f$   cross
sections, including the QED continuum background as  well as the Higgs
contribution, and introducing  CP-conserving and CP-violating observables
in polarized $\gamma \gamma$  collisions. In this connection,  we also
review  the  formalism for three-way  Higgs  mixing, stressing the key
role played by off-diagonal absorptive parts. Section~3 presents model
calculations of cross sections and CP-violating asymmetries in $\tau^+
\tau^-$ and  ${\bar  b} b$ final states  in  a  specific scenario with
strong three-way neutral-Higgs  mixing. Section~4 discusses ${\bar
t} t$ final states in a scenario with two strongly-mixed heavy neutral
Higgs  bosons.  Finally, Section~5   summarizes  our conclusions   and
presents some prospects for future work.

\setcounter{equation}{0}
\section{Polarization-Dependent Cross Sections\newline in $\gamma
\gamma$ Collisions}
\label{sec:crosssections}

We study in this Section the processes $\gamma\gamma\rightarrow f
\bar{f}$, where $f=\mu^-, \tau^-, b$, or $t$.  We consider three separate 
cases
for the helicities of the initial-state photons and the final-state
fermions, and present a classification of all the polarization-dependent
cross sections according to their CP and CP$\widetilde{\rm T}$ parities.

\subsection{QED Continuum Background}
\label{sec:continuum}

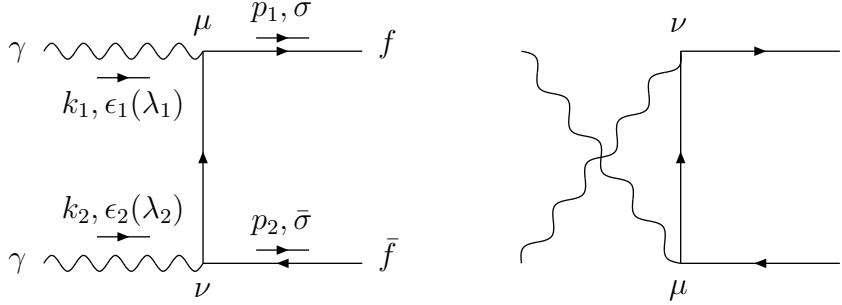
\begin{figure}[tbh]
\vspace*{1cm}
\begin{center}
\begin{picture}(300,100)(0,0)

\Photon(0,90)(60,90){3}{5}
\Photon(0,10)(60,10){3}{5}
\ArrowLine(60,90)(120,90)
\ArrowLine(60,10)(60,90)
\ArrowLine(120,10)(60,10)

\Text(-10,90)[]{$\gamma$}
\Text(60,100)[]{$\mu$}
\ArrowLine(20,80)(40,80)
\Text(30,70)[]{$k_1,\epsilon_1(\lambda_1)$}

\Text(-10,10)[]{$\gamma$}
\Text(60,0)[]{$\nu$}
\ArrowLine(20,20)(40,20)
\Text(30,30)[]{$k_2,\epsilon_2(\lambda_2)$}

\Text(130,93)[]{$f$}
\ArrowLine(80,95)(100,95)
\Text(90,105)[]{$p_1,\sigma$}

\Text(130,13)[]{$\bar{f}$}
\ArrowLine(80,15)(100,15)
\Text(90,25)[]{$p_2,\bar{\sigma}$}

\Photon(180,10)(240,90){3}{5}
\Photon(180,90)(240,10){3}{5}
\ArrowLine(240,90)(300,90)
\ArrowLine(240,10)(240,90)
\ArrowLine(300,10)(240,10)

\Text(240,100)[]{$\nu$}
\Text(240,0)[]{$\mu$}

\end{picture}\\
\end{center}
\smallskip
\noindent
\caption{\it Feynman diagrams contributing to the tree-level QED 
background, introducing our definitions of the initial-state photon and 
final-state fermion momenta and helicities.}
\label{fig:QED}
\end{figure}

The tree-level Feynman diagrams for the QED process $\gamma \gamma 
\rightarrow {\bar f} f$ are shown in Fig.~\ref{fig:QED}.
In the two-photon centre-of-mass (c.o.m.) system, the helicity amplitudes 
for the QED production of a fermion-antifermion pair take the forms:
\begin{eqnarray}
{\cal M}_C &=& 4\pi\alpha Q_f^2  \
\langle \sigma\,\bar{\sigma};\lambda_1\,\lambda_2 \rangle_C ,
\label{eq:MC}
\end{eqnarray}
where
\begin{eqnarray}
\langle \sigma\,\sigma;\lambda\,\lambda \rangle_C &=& 
\frac{4 m_f}{\sqrt{\hat{s}}}\ 
\frac{1}{1-\beta_f^2 c_\theta^2} \ (\lambda+\sigma \beta_f)\,,
\nonumber \\
\langle \sigma\,\sigma;\lambda\,-\lambda \rangle_C &=& 
-\frac{4 m_f}{\sqrt{\hat{s}}}\ 
\frac{s_\theta^2}{1-\beta_f^2 c_\theta^2} \ \sigma \beta_f\,,
\nonumber \\
\langle \sigma\,-\sigma;\lambda\,\lambda \rangle_C &=& 0\,,
\nonumber \\
\langle \sigma\,-\sigma;\lambda\,-\lambda \rangle_C &=& 
-2 \beta_f \ 
\frac{s_\theta}{1-\beta_f^2 c_\theta^2} \ (\sigma\lambda+c_\theta) \,,
\label{eq:QEDhel}
\end{eqnarray}
where $s_\theta\equiv \sin\theta$ and $c_\theta\equiv \cos\theta$
with $\theta$ the angle between ${\bf p_1}$ and ${\bf k_1}$,
and $\beta_f \equiv \sqrt{1-4m_f^2/\hat{s}}$ with 
$\hat{s} = (k_1+k_2)^2 = (p_1+p_2)^2$.
We allow for independent and measurable polarizations $\lambda_{1,2}$ of the 
initial-state photons and ${\bar \sigma}, \sigma$ of
the final-state fermion-antifermion pair. We note that the last amplitude 
in (\ref{eq:QEDhel}) with completely different
helicity states is the least important, since the Higgs-mediated diagram 
is non-vanishing only when the helicities of photons and/or those of 
final fermions are equal.


\subsection{Coupled-Channel Analysis of Processes Mediated\newline 
by Higgs  Bosons}

We now discuss how mixed Higgs bosons may contribute to the various 
$\gamma \gamma \rightarrow {\bar f} f$ helicity amplitudes, interfering 
with the above QED amplitudes and, in some cases, violating either CP 
and/or invariance under the CP$\widetilde{\rm T}$ transformation defined 
below.

In situations where two or more MSSM Higgs bosons contribute
simultaneously to the production of some fermion-antifermion pair, one
should consider~\cite{ELP1}\footnote{For an alternative approach, see
Ref.~\cite{CKLZ}.} the `full' $3 \times 3$ Higgs-boson propagator
matrix $D(\hat{s})$, including off-diagonal absorptive parts~\footnote{As
commented in~\cite{ELP1}, the complete propagator matrix $D(\hat{s})$ is a
$4\times 4$-dimensional matrix spanned by the
basis~$(H_1,H_2,H_3,G^0)$~\cite{APNPB}. However, the small off-resonant
self-energy transitions of the Higgs bosons $H_{1,2,3}$ to the neutral
would-be Goldstone boson $G^0$ may safely be neglected for our purposes.}.
This is given by
\begin{equation}
  \label{eq:hprop}
D (\hat{s}) = \hat{s}\,
\left(\begin{array}{ccc}
       \hat{s}-M_{H_1}^2+i\imag\widehat{\Pi}_{11}(\hat{s}) &
i\imag\widehat{\Pi}_{12}(\hat{s})&
       i\imag\widehat{\Pi}_{13}(\hat{s}) \\
       i\imag\widehat{\Pi}_{21}(\hat{s}) &
\hat{s}-M_{H_2}^2+i\imag\widehat{\Pi}_{22}(\hat{s})&
       i\imag\widehat{\Pi}_{23}(\hat{s}) \\
       i\imag\widehat{\Pi}_{31}(\hat{s}) & 
i\imag\widehat{\Pi}_{32}(\hat{s}) &
       \hat{s}-M_{H_3}^2+
       i\imag\widehat{\Pi}_{33}(\hat{s})
      \end{array}\right)^{-1} ,
\label{eq:Hprop}
\end{equation}
where $M_{H_{1,2,3}}$ are the one-loop Higgs-boson pole masses, and
higher-order absorptive effects on $M_{H_{1,2,3}}$ have been
ignored~\cite{CEPW2}.The absorptive part of the Higgs-boson propagator
matrix receives contributions from loops of fermions, vector bosons,
associated pairs of Higgs and vector bosons, Higgs-boson pairs, and
sfermions: 
\begin{equation} 
\imag\widehat{\Pi}_{ij}(s)=
\imag\widehat{\Pi}^{ff}_{ij}(s)+
\imag\widehat{\Pi}^{VV}_{ij}(s)+\imag\widehat{\Pi}^{HV}_{ij}(s) +
\imag\widehat{\Pi}^{HH}_{ij}(s) +
\imag\widehat{\Pi}^{\tilde{f}\tilde{f}}_{ij}(s)\,, 
\end{equation}
respectively. We use the pinch technique (PT)~\cite{PINCH1} to evaluate the
contributions from pairs of vector bosons and associated pairs of Higgs
and vector bosons, following the procedure used in~\cite{PINCH2} for the SM
Higgs sector.


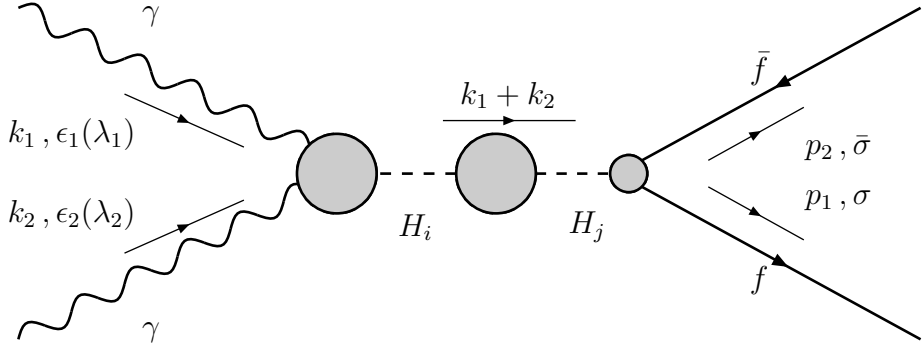
\begin{figure}[tbh]
\vspace*{1cm}
\begin{center}

\SetWidth{1.0}

\begin{picture}(250,100)(0,0)
\Photon(-50,112.5)(70,55){4}{6}
\Photon(-50,-12.5)(70,50){4}{6}
\DashCArc(130,50)(15,0,360){4}
\GOval(130,50)(15,15)(0){0.8}
\DashLine(70,50)(115,50){4}
\DashCArc(70,50)(15,0,360){4}
\GOval(70,50)(15,15)(0){0.8}
\DashLine(145,50)(180,50){4}
\DashCArc(180,50)(7,0,360){4}
\GOval(180,50)(7,7)(0){0.8}
\ArrowLine(290,112.5)(185,55)
\ArrowLine(185,45)(290,-12.5)
\Text(0,110)[]{$\gamma$}
\Text(0,-10)[]{$\gamma$}
\Text(100,30)[]{$H_i$}
\Text(165,30)[]{$H_j$}
\Text(230,90)[]{$\bar{f}$}
\Text(230,10)[]{$f$}

\SetWidth{0.5}
\ArrowLine(210,55)(246,75)
\ArrowLine(210,45)(246,25)
\Text(260,60)[]{$p_2\,,\bar{\sigma}$}
\Text(260,40)[]{$p_1\,,\sigma$}
\ArrowLine(-10,20)(35,40)
\ArrowLine(-10,80)(35,60)
\Text(-30,35)[]{$k_2\,,\epsilon_2 (\lambda_2)$}
\Text(-30,65)[]{$k_1\,,\epsilon_1 (\lambda_1)$}
\ArrowLine(110,70)(160,70)
\Text(135,80)[]{$k_1+k_2$}
\end{picture}
\end{center}
\smallskip
\noindent
\caption{\it Mechanisms contributing to the process $\gamma \gamma
\rightarrow H_{1,2,3} \to {\bar f} f$, including off-diagonal
absorptive parts in the Higgs-boson propagator matrix.}
\label{fig:Higgs}
\end{figure}


The helicity amplitudes contributing to 
$\gamma \gamma \rightarrow H \rightarrow {\bar f} f$, see
Fig.~\ref{fig:Higgs}, are given by
\begin{equation}
{\cal M}_H \; = \; \frac{\alpha \, m_f \, \sqrt{\hat{s}}}{4\pi v^2}
\langle \sigma ;\lambda_1 \rangle_H 
\delta_{\sigma\bar{\sigma}}\delta_{\lambda_1\lambda_2},
\label{eq:MH}
\end{equation}
where the reduced amplitude
\begin{equation}
\langle \sigma ;\lambda \rangle_H \; = \; \sum_{i,j=1}^3
[S_i^\gamma(\sqrt{\hat{s}})+i\lambda P_i^\gamma(\sqrt{\hat{s}})]
\ D_{ij}(\hat{s}) \ 
(\sigma\beta_f g^S_{H_j\bar{f}f}-ig^P_{H_j\bar{f}f}),
\label{eq:redHamp}
\end{equation}
is a quantity given by the Higgs-boson propagator matrix (\ref{eq:Hprop})
combined with the 
production and decay vertices.  The one-loop induced complex couplings of
the $\gamma \gamma H_i$ vertex,
$S_i^\gamma(\sqrt{\hat{s}})$ and $P_i^\gamma(\sqrt{\hat{s}})$,
get dominant contributions from charged particles
such as the bottom and top quarks, tau leptons, $W^\pm$ bosons,
charginos, third-generation sfermions and charged Higgs bosons.
Relevant aspects of the loop-induced
corrections to the $H_jf \bar{f}$ vertices were discussed
in~\cite{ELP1}. We follow the convention   of {\tt
CPsuperH}~\cite{CPsuperH} for the couplings   of the Higgs bosons.


For future reference, we note the following properties of the $\gamma 
\gamma
\rightarrow
{\bar f} f$ helicity amplitudes under the CP transformation:
\begin{equation}
\langle \sigma\,\bar{\sigma};\lambda_1\,\lambda_2 \rangle  \ \
\stackrel{\rm CP}{\leftrightarrow} \ \
(-1)(-1)^{(\sigma-\bar{\sigma})/2}
\langle -\bar{\sigma}\,-\sigma;-\lambda_2\,-\lambda_1 \rangle  .
\label{eq:cp}
\end{equation}
Also interesting are the properties under the CP$\widetilde{\rm T}$
transformation, where $\widetilde{\rm T}$ reverses the signs of the spins 
and the three-momenta of the
asymptotic states, without interchanging initial and final
states, and the matrix element gets complex conjugated:
\begin{equation}
\langle \sigma\,\bar{\sigma};\lambda_1\,\lambda_2 \rangle  \ \
\stackrel{\rm CP\widetilde{\rm T}}{\leftrightarrow} \ \
(-1)(-1)^{(\sigma-\bar{\sigma})/2}
\langle -\bar{\sigma}\,-\sigma;-\lambda_2\,-\lambda_1 \rangle^*  .
\label{eq:cpt}
\end{equation}
Evidently, the QED helicity amplitudes (\ref{eq:QEDhel}) are even under
both the CP and CP$\widetilde{\rm T}$ transformations. 
On the other hand, the simultaneous presence of 
$\{S_i^\gamma, P_i^\gamma\}$ and/or $\{g^S_{H_j\bar{f}f}, g^P_{H_j\bar{f}f}\}$
would signal CP violation in the Higgs-boson-exchange amplitude
(\ref{eq:redHamp}), and non-vanishing absorptive parts from the vertices 
and the propagators could also lead to 
CP$\widetilde{\rm T}$ violation in the Higgs-exchange diagram. 

It is convenient to distinguish three distinct cases for the helicities of 
the initial-state photons and final-state fermions. 

\subsection{Case I: Identical photon and fermion helicities}

The first case is 
that of identical photon and fermion helicities, in which case the 
amplitude may be written as
\begin{eqnarray}
{\cal M}^{\rm I}_{\sigma\lambda}=
\left. {\cal M}_C \right|_{\bar{\sigma}=\sigma, \lambda_1=\lambda_2=\lambda}
+{\cal M}_H
&=& \frac{\alpha \, m_f \, \sqrt{\hat{s}}}{4\pi v^2}
\langle \sigma ;\lambda \rangle
\end{eqnarray}
where ${\cal M}_C$ was given in (\ref{eq:MC}), ${\cal M}_H$ was given in 
(\ref{eq:MH}) and we have
\begin{equation}
\langle \sigma ;\lambda \rangle \; \equiv \;
 \langle \sigma ;\lambda \rangle_H + R(\hat{s}) f(\theta)
\langle \sigma ;\lambda \rangle_C \,.
\label{eq:helamp}
\end{equation}
The previous discussion of the QED amplitude ${\cal M}_C$ (\ref{eq:MC}) 
yields
\begin{eqnarray}
R(\hat{s}) &=& 64\pi^2 Q_f^2\, v^2/\hat{s}\,, \nonumber \\
f(\theta) &=& 1/(1-\beta_f^2 c_\theta^2)\,, \nonumber  \\
\langle \sigma ;\lambda \rangle_C  &=& \lambda+\sigma\beta_f \,,
\label{eq:slC}
\end{eqnarray}
and we note that $\langle \pm ;\mp \rangle_C  = \mp(1-\beta_f)$, which 
vanishes in the limit $m_f \rightarrow 0$.

Corresponding to the different combinations of helicities of the 
initial-state photons and final-state fermions, we have four cross 
sections:
\begin{equation}
\frac{d\hat{\sigma}_{\sigma\lambda}}{dc_\theta}
=\frac{\beta_f N_C}{32 \pi \hat{s}}
\left|{\cal M}^{\rm I}_{\sigma\lambda}\right|^2  
\end{equation}
where $\sigma\,,\lambda=\pm\,.$ After integrating over $c_\theta$, we have
\begin{equation}
\hat{\sigma}_{\sigma\lambda}
=\frac{\beta_f N_C}{32 \pi}
\left(\frac{\alpha \, m_f }{4\pi v^2} \right)^2
{\cal Y}_{\sigma\lambda}
\end{equation}
with 
\begin{eqnarray}
{\cal Y}_{\sigma\lambda}&\equiv &
\int_{-z_f}^{z_f} dc_\theta \left|\langle\sigma;\lambda\rangle\right|^2=
2\left|\langle \sigma;\lambda \rangle_H \right|^2
+2R(\hat{s})^2\, F_1^{z_f} \left|\langle \sigma;\lambda \rangle_C \right|^2
\nonumber \\
&&\hspace{3cm}
+2R(\hat{s})\, F_2^{z_f} \ \real(\langle \sigma;\lambda \rangle_H
\langle \sigma;\lambda \rangle_C^*) \,. 
\label{eq:y}
\end{eqnarray}
The functions $F_1^{z_f}$ and $F_2^{z_f}$ are given by
\begin{eqnarray}
F_1^{z_f} &=&\frac{1}{2}\int_{-z_f}^{z_f}dc_\theta f^2(\theta)=
\frac{z_f}{2(1-z_f^2\beta_f^2)}+ \frac{\ln{\frac{1+z_f\beta_f}{1-z_f\beta_f}}}{4\beta_f}\,,
\nonumber \\
F_2^{z_f} &=&\int_{-z_f}^{z_f}dc_\theta f(\theta)
=\frac{\ln{\frac{1+z_f\beta_f}{1-z_f\beta_f}}}{\beta_f} \,.
\label{eq:f12}
\end{eqnarray} 
Note that we have introduce an experimental cut on the fermion polar angle 
$\theta$ : $|\cos\theta| \leq z_f$ and $\cos\theta^f_{\rm cut}=z_f$. 
Experimentally, we can not measure the final state
fermion if it has too small angle $\theta$ outside the coverage of detectors.
This angular cut has significant effects in the cases of light fermions, 
$f=\mu$, $\tau$, and $b$, since the QED continuum cross section 
$\hat{\sigma}_C$, or $F_1^{z_f}$, strongly depends on it.  
Actually we find that the QED cross sections are suppressed by
factors of about 5000 and 20 for $f=\mu$ and $f=\tau$ cases, respectively, 
by imposing $\theta^{\mu,\tau}_{\rm cut}=130$ mrad angle cut 
($z_{\mu,\tau}\simeq 0.99$) when $\sqrt{\hat{s}}=120$ GeV.
For $b$-quark case, the suppression factor is about 30 imposing 
$\theta^b_{\rm cut}=280$ mrad ($z_b\simeq 0.96$).
On the other hand, the Higgs-mediated cross section and the QED continuum 
cross section for top quarks are hardly affected by the polar angle cut.
The introduction on the polar angle cut, therefore, greatly enhance the
significance of the Higgs-mediated process with respect to the QED continuum
one for $f=\mu\,,\tau\,,$ and $b$.

The helicity-averaged cross section is
\begin{equation}
\hat{\sigma}=\frac{1}{4}\left(
\hat{\sigma}_{++}+\hat{\sigma}_{--}+
\hat{\sigma}_{+-}+\hat{\sigma}_{-+} \right) \,.
\end{equation}
We can construct two CP-violating cross sections in 
terms of $\hat{\sigma}_{\sigma\lambda}$:
\begin{equation}
\hat\Delta_1 \equiv \hat{\sigma}_{++}-\hat{\sigma}_{--}\,, \ \
\hat\Delta_2 \equiv \hat{\sigma}_{+-}-\hat{\sigma}_{-+}\,,
\end{equation}
or, equivalently, the two linear combinations
\begin{equation}
(\hat\Delta_1  + \hat\Delta_2) = \sum_{\lambda=\pm}
(\hat{\sigma}_{+\lambda}-\hat{\sigma}_{-\lambda})\,, \ \
(\hat\Delta_1  - \hat\Delta_2) = \sum_{\sigma=\pm}
(\hat{\sigma}_{\sigma +}-\hat{\sigma}_{\sigma -})\,.
\end{equation}
The CP-violating cross section ($\hat\Delta_1-\hat\Delta_2$) can be 
measured without determining the helicities of the final-state fermions.

Finally, the QED continuum contribution is given by
\begin{equation}
\hat{\sigma}_C=\frac{\beta_f N_C}{16 \pi}
\left(\frac{\alpha \, m_f }{4\pi v^2} \right)^2
\frac{R(\hat{s})^2F^{z_f}_1
\sum_{\sigma,\lambda=\pm}
\left|\langle\sigma;\lambda\rangle_C\right|^2}{4}\,.
\end{equation}
Note that, when $z_{\mu\,,\tau\,,b}=1$, the leading term of
$\hat{\sigma}_C$ is proportional to $N_C\,Q_f^4$ with a factor $m_f^2$
cancelled.  But with $z_f\lsim 0.99$, 
$F_1^{z_f}$ becomes nearly independent of fermion species and the cross
section is proportional to $N_C\,m_f^2\,Q_f^4$.

\subsection{Case II: Identical fermion helicities}

In this case, there are two possible final-state fermion helicity states 
to consider, so the amplitude becomes a matrix:
\begin{equation}
\left({\cal M}^{\rm II}_\sigma\right)_{\lambda_1\lambda_2}=
\frac{\alpha \, m_f \, \sqrt{\hat{s}}}{4\pi v^2} \left(
\begin{array}{cc}
\langle \sigma;+ \rangle & \langle\sigma\rangle_C  \\
\langle\sigma\rangle_C & \langle \sigma;- \rangle 
\end{array} \right)
\end{equation}
where $\sigma=\pm$,
\begin{equation}
\langle \sigma \rangle_C \equiv -R(\hat{s})\,s_\theta^2\,f(\theta)\, 
\sigma\beta_f \,,
\end{equation}
and the
$\langle \sigma;\lambda \rangle $ were defined in (\ref{eq:helamp}).

The polarization density matrices for the two photons are:
\begin{eqnarray}
\tilde{\rho}=\frac{1}{2}\left(\begin{array}{cc}
                   1+\tilde{\zeta}_2 & -\tilde{\zeta}_3+i\tilde{\zeta}_1 \\
                   -\tilde{\zeta}_3-i\tilde{\zeta}_1 & 1-\tilde{\zeta}_2\\
                   \end{array}\right) \,, \ \
\rho=\frac{1}{2}\left(\begin{array}{cc}
                   1+\zeta_2 & -\zeta_3+i\zeta_1 \\
                   -\zeta_3-i\zeta_1 & 1-\zeta_2\\
                   \end{array}\right) \,, \ \
\end{eqnarray}
where the $\zeta_i$ ($\tilde{\zeta}_i$) are the Stokes parameters 
which describe the polarization transfer from the initial
laser light and electron (positron)
to the colliding photon: $\zeta_2$ is the degree of circular 
polarization and $(\zeta_3,\zeta_1)$ are the degrees of linear polarization 
transverse and normal to the plane defined by the electron direction and 
the direction of the maximal linear polarization of the initial laser light. 
The polarization-weighted squared matrix elements can be obtained by
~\cite{HaZe}
\begin{equation}
\left|{\cal M}^{\rm II}_\sigma\right|^2
={\rm Tr} \left[ {\cal M}^{\rm II}_\sigma \, \tilde{\rho} \,
 {\cal M}^{\rm II\,\dagger}_\sigma \, \rho^T \right] \,.
\end{equation}
The initial-spin average factor has
already been included in the spin density matrices, so that summing over 
$\zeta_i, \tilde{\zeta}_i$ and $\sigma$ gives the total cross section. 
The amplitude squared can therefore be written as
\begin{eqnarray}
\left|{\cal M}^{\rm II}_\sigma\right|^2 &=& 
\left(\frac{\alpha m_f \sqrt{\hat{s}}}{4\pi v^2}\right)^2  \nonumber \\
&\times &
\Bigg\{ 
A^\sigma_1(1+\zeta_2\tilde{\zeta}_2)+
A^\sigma_2[(\zeta_1\tilde{\zeta}_1-\zeta_2\tilde{\zeta}_2)+
(\zeta_3\tilde{\zeta}_3-\zeta_2\tilde{\zeta}_2)]
\nonumber \\ &&
+B^\sigma_1(\zeta_2+\tilde{\zeta}_2)
+B^\sigma_2(\zeta_1\tilde{\zeta}_3+\zeta_3\tilde{\zeta}_1)
+B^\sigma_3(\zeta_3\tilde{\zeta}_3-\zeta_1\tilde{\zeta}_1)
\nonumber \\ &&
+C^\sigma_1(\zeta_1+\tilde{\zeta}_1)
+C^\sigma_2(\zeta_3+\tilde{\zeta}_3)
+C^\sigma_3(\zeta_1\tilde{\zeta}_2+\zeta_2\tilde{\zeta}_1)
+C^\sigma_4(\zeta_2\tilde{\zeta}_3+\zeta_3\tilde{\zeta}_2)
\Bigg\} \,, \nonumber
\end{eqnarray}
where
\begin{eqnarray}
A^\sigma_1 &=& \frac{1}{4}\left(|\langle \sigma;+ \rangle|^2 +|\langle
\sigma;- \rangle|^2 
+2\,|\langle \sigma \rangle_C|^2\right) \,,
\nonumber \\
A^\sigma_2 &=& \frac{1}{2}\,|\langle \sigma \rangle_C|^2\,, \nonumber \\
B^\sigma_1 &=& \frac{1}{4}\left(|\langle \sigma;+ \rangle|^2 -|\langle
\sigma;- \rangle|^2\right) \,,
\nonumber \\
B^\sigma_2 &=& \frac{1}{2}\,\imag[\langle \sigma;+ \rangle\,\langle \sigma;- \rangle^*]
\,, \nonumber \\
B^\sigma_3 &=& \frac{1}{2}\,\real[\langle \sigma;+ \rangle\,\langle \sigma;- \rangle^*]
\,, \nonumber \\
C^\sigma_1 &=& -\frac{1}{2}\,\imag[(\langle \sigma;+ \rangle - 
\langle \sigma;- \rangle)\,
\langle \sigma \rangle_C^*] \,, \nonumber \\
C^\sigma_2 &=& -\frac{1}{2}\,\real[(\langle \sigma;+ \rangle + 
\langle \sigma;- \rangle)\,
\langle \sigma \rangle_C^*] \,, \nonumber \\
C^\sigma_3 &=& -\frac{1}{2}\,\imag[(\langle \sigma;+ \rangle + 
\langle \sigma;- \rangle)\,
\langle \sigma \rangle_C^*] \,, \nonumber \\
C^\sigma_4 &=& -\frac{1}{2}\,\real[(\langle \sigma;+ \rangle - 
\langle \sigma;- \rangle)\,
\langle \sigma \rangle_C^*] \,.
\end{eqnarray}
We note that the quantities $B^\sigma_{2,3}$ are related to the 
observables in the
interference between the amplitudes with different photon helicities requiring
linear polarizations of the colliding photon beams, and that
the observables $C^\sigma_i$ are due to interference with the QED 
continuum.

We distinguish two categories of cross sections, according to the 
final-state fermion helicity $\sigma$:
\begin{eqnarray}
\frac{d\hat\Sigma^X}{dc_\theta} &\equiv & 
\frac{\beta_f N_C}{32 \pi } 
\left(\frac{\alpha m_f }{4\pi v^2}\right)^2 (X^++X^-)\,,
\nonumber \\
\frac{d\hat\Delta^X}{dc_\theta} &\equiv & 
\frac{\beta_f N_C}{32 \pi} 
\left(\frac{\alpha m_f}{4\pi v^2}\right)^2 (X^+-X^-)\,,
\end{eqnarray}
where $X=A_i,B_j,C_k$. 
We define the following quantities obtained by integration 
over the angle $\theta$:
\begin{eqnarray}
{\cal W}_\sigma &\equiv &
\int_{-z_f}^{z_f} dc_\theta \left|\langle \sigma \rangle_C \right|^2 =
\beta_f^2 R(\hat{s})^2 F^{z_f}_3  \,, \nonumber \\
{\cal X}_\sigma &\equiv &
\int_{-z_f}^{z_f} dc_\theta \langle \sigma;+ \rangle \langle \sigma;- \rangle^* =
2\langle \sigma;+ \rangle_H \langle \sigma;- \rangle_H^*
+2R(\hat{s})^2\, F^{z_f}_1 \langle \sigma;+ \rangle_C \langle \sigma;- \rangle_C^*
\nonumber \\
&& \hspace{0 cm}
+R(\hat{s})\, F^{z_f}_2 \ 
(\langle \sigma;+ \rangle_C \langle \sigma;- \rangle_H^*
+\langle \sigma;+ \rangle_H \langle \sigma;- \rangle_C^*)
\,, \nonumber \\
%
%
{\cal Z}_{\sigma\lambda}&\equiv &
\int_{-z_f}^{z_f} dc_\theta \langle \sigma;\lambda \rangle \langle \sigma \rangle_C^*
= -\sigma\beta_f R(\hat{s})\left[
F^{z_f}_4 \langle \sigma;\lambda \rangle_H +
R(\hat{s}) F^{z_f}_5 \langle \sigma;\lambda \rangle_C \right] \,,
\end{eqnarray}
and define ${\cal Y}_{\sigma\lambda}$ as in (\ref{eq:y}). In addition to the
functions $F_{1,2}^{z_f}$ given in (\ref{eq:f12}), the functions
$F_{3,4,5}^{z_f}$ are
\begin{eqnarray}
F^{z_f}_3&=&\int_{-z_f}^{z_f} dc_\theta s_\theta^4 f^2(\theta)
=\frac{[3-2(1+z_f^2)\beta_f^2+\beta_f^4]\,z_f}{\beta_f^4\,(1-z_f^2\beta_f^2)}
-\frac{\ln{\frac{1+z_f\beta_f}{1-z_f\beta_f}}}{2\beta_f^5}(1-\beta_f^2)(3+\beta_f^2)
\,, \nonumber \\
F^{z_f}_4&=&\int_{-z_f}^{z_f} dc_\theta s_\theta^2 f(\theta) = \frac{2\,z_f}{\beta_f^2}
-\frac{\ln{\frac{1+z_f\beta_f}{1-z_f\beta_f}}}{\beta_f^3}(1-\beta_f^2)
\,, \nonumber \\
F^{z_f}_5&=&\int_{-z_f}^{z_f} dc_\theta s_\theta^2 f^2(\theta) =
\frac{-(1-\beta_f^2)\,z_f}{\beta_f^2\,(1-z_f^2\beta_f^2)}
+\frac{\ln{\frac{1+z_f\beta_f}{1-z_f\beta_f}}}{2\beta_f^3}(1+\beta_f^2)
\,. \nonumber 
\end{eqnarray}
Then, all the 18 cross sections $\hat\Sigma^X$ and $\hat\Delta^X$ can
be written in terms of ${\cal W}_\sigma$,
${\cal X}_\sigma$, ${\cal Y}_{\sigma\lambda}$, and ${\cal Z}_{\sigma\lambda}$.
For example, the total cross section $\hat\Sigma^{A_1}$ is
\begin{equation}
\hat\Sigma^{A_1}= 
\frac{\beta_f N_C}{32 \pi } 
\left(\frac{\alpha m_f }{4\pi v^2}\right)^2
\frac{({\cal Y}_{++}+{\cal Y}_{+-}+2{\cal W}_{+})
+({\cal Y}_{-+}+{\cal Y}_{--}+2{\cal W}_{-})}{4} \,,
\end{equation}
and the other analogous quantities are given in the second column of 
Table~\ref{tab:AllCases}.

The CP and CP$\tilde{\rm T}$ parities of the
various cross sections are also shown in the first column of
Table \ref{tab:AllCases}.
The CP and CP$\widetilde{\rm T}$ parities of the polarization-dependent cross
sections can easily be obtained by observing that
\begin{eqnarray}
&&
{\cal W}_\sigma  \stackrel{\rm CP}{\leftrightarrow} \ {\cal W}_{-\sigma}\,, \ \
{\cal X}_\sigma \ \stackrel{\rm CP}{\leftrightarrow} \ {\cal X}^*_{-\sigma}\,, \ \
{\cal Y}_{\sigma\lambda} \ \stackrel{\rm CP}{\leftrightarrow} 
\ {\cal Y}_{-\sigma-\lambda}\,, \ \
{\cal Z}_{\sigma\lambda}  \stackrel{\rm CP}{\leftrightarrow} 
\ {\cal Z}_{-\sigma-\lambda}\,, \nonumber \\
&&
{\cal W}_\sigma \stackrel{\rm CP\widetilde{\rm T}}{\leftrightarrow} 
{\cal W}_{-\sigma}\,, \ \
{\cal X}_\sigma \stackrel{\rm CP\widetilde{\rm T}}{\leftrightarrow} 
{\cal X}_{-\sigma}\,, \ \
{\cal Y}_{\sigma\lambda} \stackrel{\rm CP\widetilde{\rm T}}{\leftrightarrow} 
{\cal Y}_{-\sigma-\lambda}\,, \ \
{\cal Z}_{\sigma\lambda} \stackrel{\rm CP\widetilde{\rm T}}{\leftrightarrow} 
\ {\cal Z}^*_{-\sigma-\lambda}\,,
\end{eqnarray}
which are derived from (\ref{eq:cp}) and (\ref{eq:cpt}). We observe that the
half of the cross sections are CP odd.
\begin{table}[\hbt]
\caption{\label{tab:AllCases}
{\it All polarization-dependent cross sections considered in units of 
$\frac{\beta_f N_C}{32 \pi } \left(\frac{\alpha m_f }{4\pi v^2}\right)^2$. The
{\rm CP} and {\rm CP}$\widetilde{\rm T}$ parities of the cross sections
are also shown in the first column. Equivalent quantities are given in the 
last column. }}
\vspace{-0.3cm}
\begin{center}
\begin{tabular}{|c|c|c|}
\hline
Cross sections & Expressions &Equivalents \\
$[{\rm CP},{\rm CP\tilde{T}}]$ & & \\
\hline\hline
$\hat{\sigma}[+,+]$ &
$({\cal Y}_{++}+{\cal Y}_{+-}+{\cal Y}_{-+}+{\cal Y}_{--})/4$ &
$\hat\Sigma^{D_1}$
\\ \hline
$\hat\Delta_1[-,-]$ &
${\cal Y}_{++}-{\cal Y}_{--}$ &
\\ \hline
$\hat\Delta_2[-,-]$ &
${\cal Y}_{+-}-{\cal Y}_{-+}$ &
\\ \hline\hline
$\hat\Sigma^{A_1}[+,+]$ &
$[({\cal Y}_{++}+{\cal Y}_{+-}+2{\cal W}_+)
+({\cal Y}_{-+}+{\cal Y}_{--}+2{\cal W}_-)]/4$ &
\\ \hline
$\hat\Sigma^{A_2}[+,+]$ &
$[{\cal W}_{+}+{\cal W}_{-}]/2$ &
$\hat\Sigma^{A_1}-\hat\Sigma^{A_2}=\hat{\sigma}$ 
\\ \hline
$\hat\Sigma^{B_1}[-,-]$ &
$[({\cal Y}_{++}-{\cal Y}_{+-})+({\cal Y}_{-+}-{\cal Y}_{--})]/4$ &
$(\hat\Delta_1-\hat\Delta_2)/4\,,\hat\Delta^{D_1}$
\\ \hline
$\hat\Sigma^{B_2}[-,+]$ &
$\imag[{\cal X}_{+}+{\cal X}_{-}]/2$ &
\\ \hline
$\hat\Sigma^{B_3}[+,+]$ &
$\real[{\cal X}_{+}+{\cal X}_{-}]/2$ &
\\ \hline
$\hat\Sigma^{C_1}[-,+]$ &
$-\imag[({\cal Z}_{++}-{\cal Z}_{+-})+({\cal Z}_{-+}-{\cal Z}_{--})]/2$ &
\\ \hline
$\hat\Sigma^{C_2}[+,+]$ &
$-\real[({\cal Z}_{++}+{\cal Z}_{+-})+({\cal Z}_{-+}+{\cal Z}_{--})]/2$ &
\\ \hline
$\hat\Sigma^{C_3}[+,-]$ &
$-\imag[({\cal Z}_{++}+{\cal Z}_{+-})+({\cal Z}_{-+}+{\cal Z}_{--})]/2$ &
\\ \hline
$\hat\Sigma^{C_4}[-,-]$ &
$-\real[({\cal Z}_{++}-{\cal Z}_{+-})+({\cal Z}_{-+}-{\cal Z}_{--})]/2$ &
\\ \hline
$\hat\Delta^{A_1}[-,-]$ &
$[({\cal Y}_{++}+{\cal Y}_{+-}+2{\cal W}_+)
-({\cal Y}_{-+}+{\cal Y}_{--}+2{\cal W}_-)]/4$ &
$(\hat\Delta_1+\hat\Delta_2)/4\,,\hat\Sigma^{D_2}$
\\ \hline
$\hat\Delta^{A_2}[-,-]$ &
$[{\cal W}_{+}-{\cal W}_{-}]/2$ &
$0$
\\ \hline
$\hat\Delta^{B_1}[+,+]$ &
$[({\cal Y}_{++}-{\cal Y}_{+-})-({\cal Y}_{-+}-{\cal Y}_{--})]/4$ &
$\hat\Delta^{D_2}$
\\ \hline
$\hat\Delta^{B_2}[+,-]$ &
$\imag[{\cal X}_{+}-{\cal X}_{-}]/2$ &
\\ \hline
$\hat\Delta^{B_3}[-,-]$ &
$\real[{\cal X}_{+}-{\cal X}_{-}]/2$ &
\\ \hline
$\hat\Delta^{C_1}[+,-]$ &
$-\imag[({\cal Z}_{++}-{\cal Z}_{+-})-({\cal Z}_{-+}-{\cal Z}_{--})]/2$ &
\\ \hline
$\hat\Delta^{C_2}[-,-]$ &
$-\real[({\cal Z}_{++}+{\cal Z}_{+-})-({\cal Z}_{-+}+{\cal Z}_{--})]/2$ &
\\ \hline
$\hat\Delta^{C_3}[-,+]$ &
$-\imag[({\cal Z}_{++}+{\cal Z}_{+-})-({\cal Z}_{-+}+{\cal Z}_{--})]/2$ &
\\ \hline
$\hat\Delta^{C_4}[+,+]$ &
$-\real[({\cal Z}_{++}-{\cal Z}_{+-})-({\cal Z}_{-+}-{\cal Z}_{--})]/2$ &
\\ \hline\hline 
$\hat\Sigma^{D_1}[+,+]$ &
$[({\cal Y}_{++}+{\cal Y}_{-+})+({\cal Y}_{+-}+{\cal Y}_{--})]/4$ &
$\hat{\sigma}$
\\ \hline
$\hat\Sigma^{D_2}[-,-]$ &
$[({\cal Y}_{++}-{\cal Y}_{-+})+({\cal Y}_{+-}-{\cal Y}_{--})]/4$ &
$(\hat\Delta_1+\hat\Delta_2)/4\,,\hat\Delta^{A_1}$
\\ \hline
$\hat\Sigma^{D_3}[+,+]$ &
$-\real[\widetilde{\cal X}_{+}+\widetilde{\cal X}_{-}]/2$ &
\\ \hline
$\hat\Sigma^{D_4}[-,+]$ &
$\imag[\widetilde{\cal X}_{+}+\widetilde{\cal X}_{-}]/2$ &
\\ \hline
$\hat\Delta^{D_1}[-,-]$ &
$[({\cal Y}_{++}+{\cal Y}_{-+})-({\cal Y}_{+-}+{\cal Y}_{--})]/4$ &
$(\hat\Delta_1-\hat\Delta_2)/4\,,\hat\Sigma^{B_1}$
\\ \hline
$\hat\Delta^{D_2}[+,+]$ &
$[({\cal Y}_{++}-{\cal Y}_{-+})-({\cal Y}_{+-}-{\cal Y}_{--})]/4$ &
$\hat\Delta^{B_1}$
\\ \hline
$\hat\Delta^{D_3}[-,-]$ &
$-\real[\widetilde{\cal X}_{+}-\widetilde{\cal X}_{-}]/2$ &
\\ \hline
$\hat\Delta^{D_4}[+,-]$ &
$\imag[\widetilde{\cal X}_{+}-\widetilde{\cal X}_{-}]/2$ &
\\ \hline
\end{tabular}
\end{center}
\end{table}

\subsection{Case III: Identical photon helicities}

In this case, there are two possible initial-state photon helicity states
to consider, so the amplitude again becomes a matrix:
\begin{equation}
\left({\cal M}^{\rm III}_\lambda\right)_{\sigma\bar{\sigma}}=
\frac{\alpha \, m_f \, \sqrt{\hat{s}}}{4\pi v^2} \left(
\begin{array}{cc}
\langle +;\lambda \rangle & 0  \\
0 & \langle -;\lambda \rangle 
\end{array} \right) \,,
\end{equation}
where $\langle \pm;\lambda \rangle $ were also defined in (\ref{eq:helamp}).

The polarization density matrices for the two final-state fermions are
\begin{eqnarray}
\bar{\rho}=\frac{1}{2}\left(\begin{array}{cc}
                   1+\bar{P}_L & -\bar{P}_T e^{i\bar{\alpha}} \\
                   -\bar{P}_T e^{-i\bar{\alpha}} & 1-\bar{P}_L\\
                   \end{array}\right) \,, \ \
\rho=\frac{1}{2}\left(\begin{array}{cc}
                   1+P_L & P_T e^{-i\alpha} \\
                   P_T e^{i\alpha} & 1-P_L\\
                   \end{array}\right) \,. \ \
\end{eqnarray}
Here, $P_L$ and $\bar{P}_L$ are the longitudinal polarizations of the
fermion $f$ and antifermion $\bar{f}$, respectively,
while $P_T$ and $\bar{P}_T$ are the degrees
of transverse polarization with $\alpha$ and $\bar{\alpha}$ being
the azimuthal angles 
with respect to the production plane.
%
The polarization-weighted squared matrix elements are~\cite{HaZe}
\begin{equation}
\left|{\cal M}^{\rm III}_\lambda\right|^2
={\rm Tr} \left[ {\cal M}^{\rm III}_\lambda \, \bar{\rho}^T \,
 {\cal M}^{\rm III\,\dagger}_\lambda \, \rho \right] \,.
\end{equation}
The total helicity-averaged cross section is obtained by summing over $P$, 
$\bar{P}$
and $\lambda$ and dividing by a factor 4 to account for the initial-state 
spin average:
\begin{eqnarray}
\left|{\cal M}^{\rm III}_\lambda\right|^2 &=& 
\left(\frac{\alpha m_f \sqrt{\hat{s}}}{4\pi v^2}\right)^2  
\Bigg\{
D^\lambda_1 (1+P_L\bar{P}_L)
+D^\lambda_2 (P_L+\bar{P}_L)
\nonumber \\
&& \hspace{3cm}
+P_T\bar{P}_T[D^\lambda_3\cos(\alpha-\bar{\alpha})
+D^\lambda_4\sin(\alpha-\bar{\alpha})]
\Bigg\} \,,
\end{eqnarray}
where
\begin{eqnarray}
D_1^\lambda &=& \frac{1}{4}
\left(|\langle +;\lambda \rangle|^2 +|\langle -;\lambda \rangle|^2\right)
\,, \nonumber \\
D_2^\lambda &=& \frac{1}{4}
\left(|\langle +;\lambda \rangle|^2 -|\langle -;\lambda \rangle|^2\right)
\,, \nonumber \\
D_3^\lambda &=& -\frac{1}{2}
\real\left(\langle +;\lambda \rangle \langle -;\lambda \rangle^*\right)
\,, \nonumber \\
D_4^\lambda &=& \frac{1}{2}
\imag\left(\langle +;\lambda \rangle \langle -;\lambda \rangle^*\right) 
\,.
\end{eqnarray}
The quantities $D^\lambda_{3,4}$ are related to the observables coming 
from the interference between the amplitudes with different fermion 
helicities, for which we need to measure the transverse polarizations of 
the final fermions.
Note that the quantity $D^\lambda_1$ can be constructed without the need 
to measure the helicities of the final fermions.

Similarly to the previous case, we define two kinds of cross sections,
corresponding to the different initial-state photon helicities $\lambda$: 
\begin{eqnarray}
\frac{d\hat\Sigma^{D_i}}{dc_\theta} &\equiv& 
\frac{\beta_f N_C}{32 \pi } 
\left(\frac{\alpha m_f }{4\pi v^2}\right)^2 (D_i^++D_i^-)\,,
\nonumber \\
\frac{d\hat\Delta^{D_i}}{dc_\theta} &\equiv& 
\frac{\beta_f N_C}{32 \pi} 
\left(\frac{\alpha m_f}{4\pi v^2}\right)^2 (D_i^+-D_i^-)\,.
\end{eqnarray}
Upon angle integration, we define
\begin{eqnarray}
\widetilde{\cal X}_\lambda &\equiv  &
\int_{-z_f}^{z_f} dc_\theta 
 \langle +;\lambda \rangle   \langle -;\lambda \rangle^* =
2\langle +;\lambda \rangle_H \langle -;\lambda \rangle^*_H
+2R(\hat{s})^2\, F^{z_f}_1 
 \langle +;\lambda \rangle_C \langle -;\lambda \rangle^*_C
\nonumber \\
&& \hspace{0 cm}
+R(\hat{s})\, F^{z_f}_2 \ 
(\langle +;\lambda \rangle_C \langle -;\lambda \rangle^*_H
+\langle +;\lambda \rangle_H \langle -;\lambda \rangle^*_C) \,,
\end{eqnarray}
which satisfies
\begin{equation}
\widetilde{\cal X}_\lambda \stackrel{\rm CP}{\leftrightarrow} 
\widetilde{\cal X}^*_{-\lambda}\,, \ \ \ \
\widetilde{\cal X}_\lambda \stackrel{\rm CP\widetilde{\rm T}}{\leftrightarrow} 
\widetilde{\cal X}_{-\lambda}\,. 
\end{equation}
Expressions for the cross sections $\hat\Sigma^{D_i}$ and $\hat\Delta^{D_i}$ 
after integrating over $c_\theta$, together with the CP and 
CP$\tilde{\rm T}$ parities of the cross sections, are given in 
Table~\ref{tab:AllCases}. We, again, notice that the half of them are CP odd.

We complete this section by noting some relations between the cross 
sections, derived from the analytic expressions given above:
\begin{eqnarray}
\hat{\sigma} &=& \hat\Sigma^{A_1} -\hat\Sigma^{A_2} = \hat\Sigma^{D_1} \,,
\nonumber \\
(\hat\Delta_1-\hat\Delta_2)/4 &=& \hat\Sigma^{B_1} = \hat\Delta^{D_1} \,, \nonumber \\
(\hat\Delta_1+\hat\Delta_2)/4 &=& \hat\Delta^{A_1} = \hat\Sigma^{D_2} \,,\nonumber \\
\hat\Delta^{B_1} &=& \hat\Delta^{D_2} \,.
\end{eqnarray}
%

\setcounter{equation}{0}
\section{CP Violation in $\gamma \gamma \to {\bar b}b\,,\mu^+\mu^-\,, 
\tau^+ \tau^-$ in a Three-Way\newline Mixing Scenario}
\label{sec:threecalx}

We now present some numerical examples of CP-violating Higgs signatures in
$\mu^+\mu^-$, $\tau^+ \tau^-$, and ${\bar b} b$ production at a $\gamma \gamma$ collider.  
As already mentioned, these signatures may be enhanced at large $\tan
\beta$, and three-way mixing is potentially important for small charged
(and hence also neutral) Higgs-boson masses. Therefore, we present in this
section some numerical analyses in a specific scenario in which all the
three neutral Higgs states mix significantly.

Explicitly, we take the following parameter set:
\begin{eqnarray}
  \label{MSSM1}
&&\tan\beta=50, \ \ M_{H^\pm}^{\rm pole}=155~~{\rm GeV},  
\nonumber \\ 
&&M_{\tilde{Q}_3} = M_{\tilde{U}_3} = M_{\tilde{D}_3} =
M_{\tilde{L}_3} = M_{\tilde{E}_3} = M_{\rm SUSY} = 0.5 ~~{\rm TeV},
\nonumber \\
&& |\mu|=0.5 ~~{\rm TeV}, \ \
|A_{t,b,\tau}|=1 ~~{\rm TeV},   \ \
|M_2|=|M_1|=0.3~~{\rm TeV}, \ \ |M_3|=1 ~~{\rm TeV},
\nonumber \\
&& 
\Phi_\mu = 0^\circ, \ \
\Phi_A=\Phi_{A_t} = \Phi_{A_b} = \Phi_{A_\tau} = 90^\circ, \ \
\Phi_1 = \Phi_2 = 0^\circ,
\end{eqnarray}
and we consider the following value for the phase of the gluino mass 
parameter $M_3$: $\Phi_3 = -10^\circ$. In this case, {\tt CPsuperH} yields 
for the masses and widths of the neutral Higgs bosons: \\
\begin{eqnarray}
&&
M_{H_1}=120.2~~{\rm GeV}, \ \
M_{H_2}=121.4~~{\rm GeV}, \ \
M_{H_3}=124.5~~{\rm GeV}, \ \
\nonumber \\  &&
\Gamma_{H_1}=1.19~~{\rm GeV}, \ \ \ \
\Gamma_{H_2}=3.42~~{\rm GeV}, \ \ \ \ \ \
\Gamma_{H_3}=3.20~~{\rm GeV}.
\end{eqnarray}
We present results for each of the above polarization cases in turn.

\subsection{CP violation in $\gamma\gamma \rightarrow b\bar{b}$}
In the process $\gamma\gamma \rightarrow b\bar{b}$, the inability of measuring
the polarization of $b$ and ${\bar b}$ quarks limits us to two observables:
the CP-even total helicity-averaged cross section $\hat{\sigma}$ and the
CP-odd cross section $\hat\Delta_1-\hat\Delta_2$.

\begin{figure}[htb]
\vspace{-1.5cm}
\centerline{\epsfig{figure=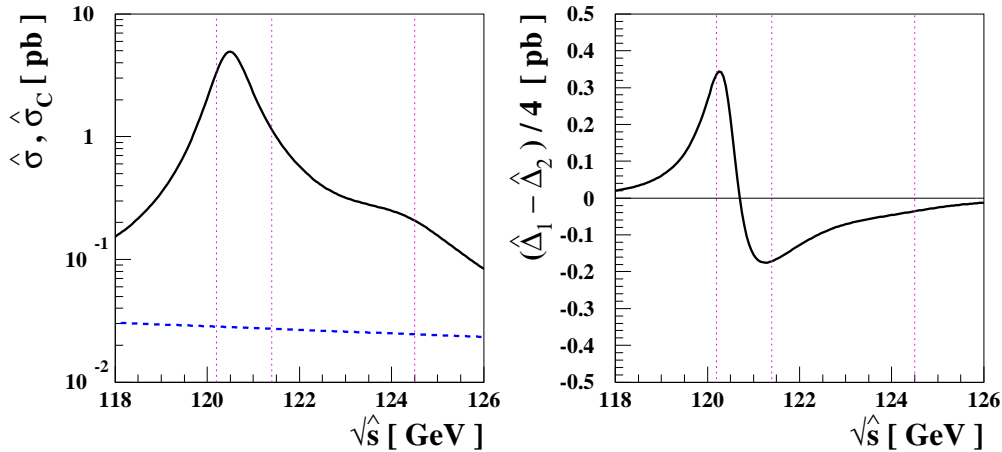,height=14cm,width=14cm}}
\vspace{-6.9 cm}
\caption{{\it 
The cross sections $\hat{\sigma}$ and $(\hat\Delta_1-\hat\Delta_2)/4$
for the process $\gamma \gamma \rightarrow {\bar b}b$ in the three-way
mixing scenario with $\Phi_3=-10^\circ$ and $\Phi_{A_{t,b}}=90^\circ$ as
functions of $\sqrt{\hat{s}}$ taking
$\theta^b_{\rm cut}=280$ mrad ($z_b\simeq 0.96$). 
The continuum cross section $\hat{\sigma}_C$ is also shown in the left frame
as a dashed line. The three Higgs masses are indicated by vertical lines. 
}}
\label{fig:ggbb}
\end{figure}

In Fig.~\ref{fig:ggbb}, we show the cross sections as functions of the
invariant mass of the bottom quarks $\sqrt{\hat s}$ taking 
$\theta^b_{\rm cut}=280$ mrad ($z_b\simeq 0.96$) coming from the coverage of
the vertex detector. In the left frame, we
also show the QED continuum cross section $\hat{\sigma}_C$ as a dashed
line. We observe the continuum contribution to the total cross section 
is negligible taking account of $\theta^b_{\rm cut}=280$ mrad.
The total helicity-averaged cross section 
is larger than about $\sim$ 0.1 pb in the region shown except around 
$\sqrt{\hat{s}}=126$ GeV.
The maximum center-of-mass energy and the luminosity in $\gamma\gamma$ 
collisions are comparable to those in $e^\pm e^-$ collisions~\cite{GKPST}.
Assuming an integrated $\gamma\gamma$ luminosity of
100 fb$^{-1}$, and high efficiency for $b$-quark reconstruction, we may
expect a sample exceeding ten thousand events. This would enable
one to probe CP asymmetry at the 1 \% level or less in the process
$\gamma\gamma \rightarrow b\bar{b}$, by controlling the polarizations of
colliding photon beams.
 
\begin{figure}[htb]
\hspace{ 4.4cm}
\centerline{\epsfig{figure=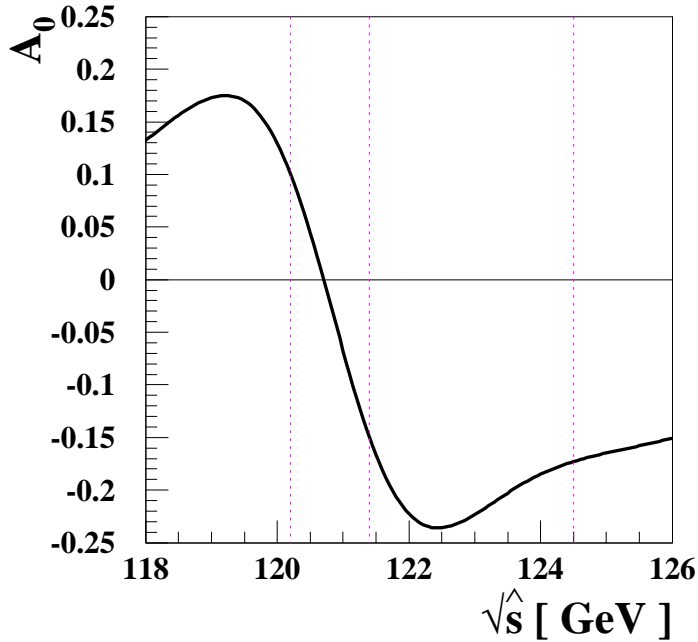,height=20cm,width=20cm}}
\vspace{-10.0 cm}
\caption{{\it 
The CP asymmetry ${\cal A}_{0}$
for the process $\gamma \gamma \rightarrow {\bar b}b$ in the three-way
mixing scenario with $\Phi_3=-10^\circ$ and $\Phi_{A_{t,b}}=90^\circ$, as
a function of $\sqrt{\hat{s}}$ taking
$\theta^b_{\rm cut}=280$ mrad ($z_b\simeq 0.96$). 
}}
\label{fig:asymbb}
\end{figure}

The CP-violating cross section $(\hat\Delta_1-\hat\Delta_2)/4$ is in a 
range between $-0.2$ pb and 0.3 pb. We define a CP asymmetry as:
\begin{equation}
{\cal A}_{0}\equiv
\frac{\hat\Delta_1- \hat\Delta_2} {4\,\hat{\sigma}}\,.
\end{equation}
In Fig.~\ref{fig:asymbb}, we show this CP asymmetry as a function of 
$\sqrt{\hat s}$. We observe that $|{\cal A}_0|$
is larger than 1 \% over most of the region
and can be as large as 25 \%. 

On the other hand, it should be noted that the $b\bar{b}$ mass resolution
in a realistic detector at a $\gamma\gamma$ collider is expected to be
several GeV~\cite{Asner}, in which case the regions of positive and
negative ${\cal A}_{0}$ would be somewhat smeared. However, at least in
this example, the integrated asymmetry should still be non-zero and
observable. Nevertheless, this physics example indicates that some premium
should be set on a detector capable of good $b\bar{b}$ mass resolution.

\subsection{CP violation in $\gamma\gamma \rightarrow \mu^+\mu^-$}

In this process, as in the process $\gamma\gamma \rightarrow \bar{b}b$, 
the inability to measure the polarization of muons limits us to two 
observables: the CP-even total helicity-averaged cross section 
$\hat{\sigma}$ and the CP-odd cross section $(\hat\Delta_1-\hat\Delta_2)/4$.
But, differently from the $\gamma\gamma \rightarrow \bar{b}b$ process, 
the good resolution in the invariant mass of the muons, which is
expected to be better than 1 GeV, enables us to examine
the $\sqrt{\hat{s}}$ dependence of the cross sections and the CP asymmetry
in the process $\gamma\gamma \rightarrow \mu^+\mu^-$.

\begin{figure}[htb]
\vspace{-1.0cm}
\centerline{\epsfig{figure=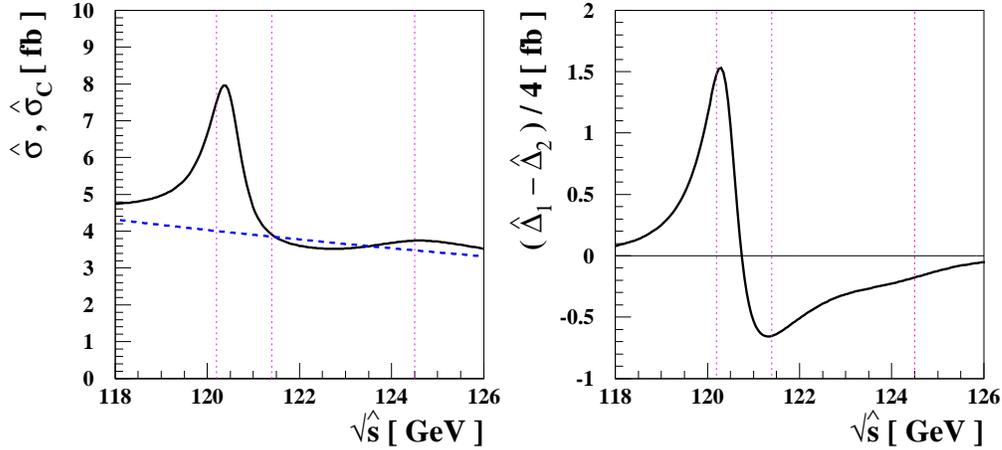,height=14cm,width=14cm}}
\vspace{-6.9 cm}
\caption{{\it 
The cross sections $\hat{\sigma}$ and $(\hat\Delta_1-\hat\Delta_2)/4$
for the process $\gamma \gamma \rightarrow \mu^+\mu^-$ in the three-way
mixing scenario with $\Phi_3=-10^\circ$ and $\Phi_{A_{t,b}}=90^\circ$, as
functions of $\sqrt{\hat{s}}$ with $\theta^\mu_{\rm cut}=130$ mrad 
($z_\mu\simeq 0.99$).
The continuum cross section $\hat{\sigma}_C$ is also shown in the left frame
as a dashed line. The three Higgs masses are indicated by vertical lines. 
}}
\label{fig:ggmm}
\end{figure}

\begin{figure}[htb]
\hspace{ 4.4cm}
\centerline{\epsfig{figure=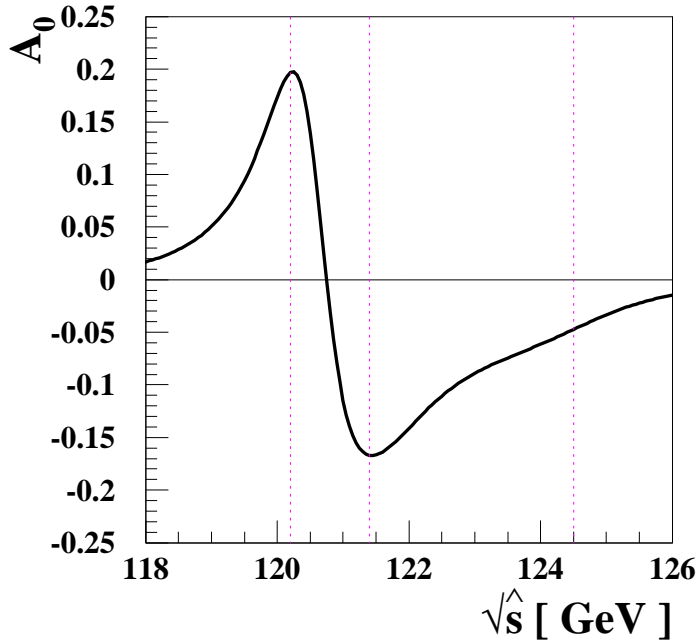,height=20cm,width=20cm}}
\vspace{-10.0 cm}
\caption{{\it 
The CP asymmetry ${\cal A}_{0}$
for the process $\gamma \gamma \rightarrow \mu^+\mu^-$ in the three-way
mixing scenario with $\Phi_3=-10^\circ$ and $\Phi_{A_{t,b}}=90^\circ$, as
a function of $\sqrt{\hat{s}}$ with $\theta^\mu_{\rm cut}=130$ mrad.
}}
\label{fig:asymmm}
\end{figure}

In Figs.~\ref{fig:ggmm} and \ref{fig:asymmm}, we show the cross sections
$\hat{\sigma}$ and $(\hat{\Delta}_1-\hat{\Delta}_2)/4$ and the CP asymmetry
${\cal A}_0$. The angle cut $\theta^\mu_{\rm cut}=130$ 
mrad, which corresponds to $z_\mu\simeq 0.99$, has been taken. 
The cross section $\hat{\sigma}$ is larger than $\sim$ 4 fb and the CP
asymmetry can be as large as 20 \%.
With 100 fb$^{-1}$ integrated $\gamma\gamma$ luminosity, 
we expect to have a sample of a few hundred events which is
enough to probe a CP asymmetry larger than 10 \%.

\subsection{CP violation in $\gamma\gamma \rightarrow \tau^+\tau^-$}

In this subsection, we consider all the observables introduced in
Sec.~\ref{sec:crosssections}, since the polarizations of
tau leptons can be measured in the process
$\gamma\gamma \rightarrow \tau^+\tau^-$ \cite{GKS}.
For the process, we impose $\theta^\tau_{\rm cut}=130$ mrad 
($z_{\tau}\simeq 0.99$) which is the
same as $\theta^\mu_{\rm cut}$. We find that the CP-even cross sections
$\hat{\sigma}$, $\hat{\Sigma}^{A_1}$, $\hat{\Delta}^{B_1}$,
$\hat{\Sigma}^{D_1}$, and $\hat{\Delta}^{D_2}$,
which are expressed in terms of ${\cal Y}_{\sigma\lambda}$, 
are very sensitive to the value of $\theta^\mu_{\rm cut}$. But the other
observables are nearly insensitive 
as long as $(1-z_\tau) \ll 1$.

\begin{figure}[htb]
\vspace{-1.5cm}
\centerline{\epsfig{figure=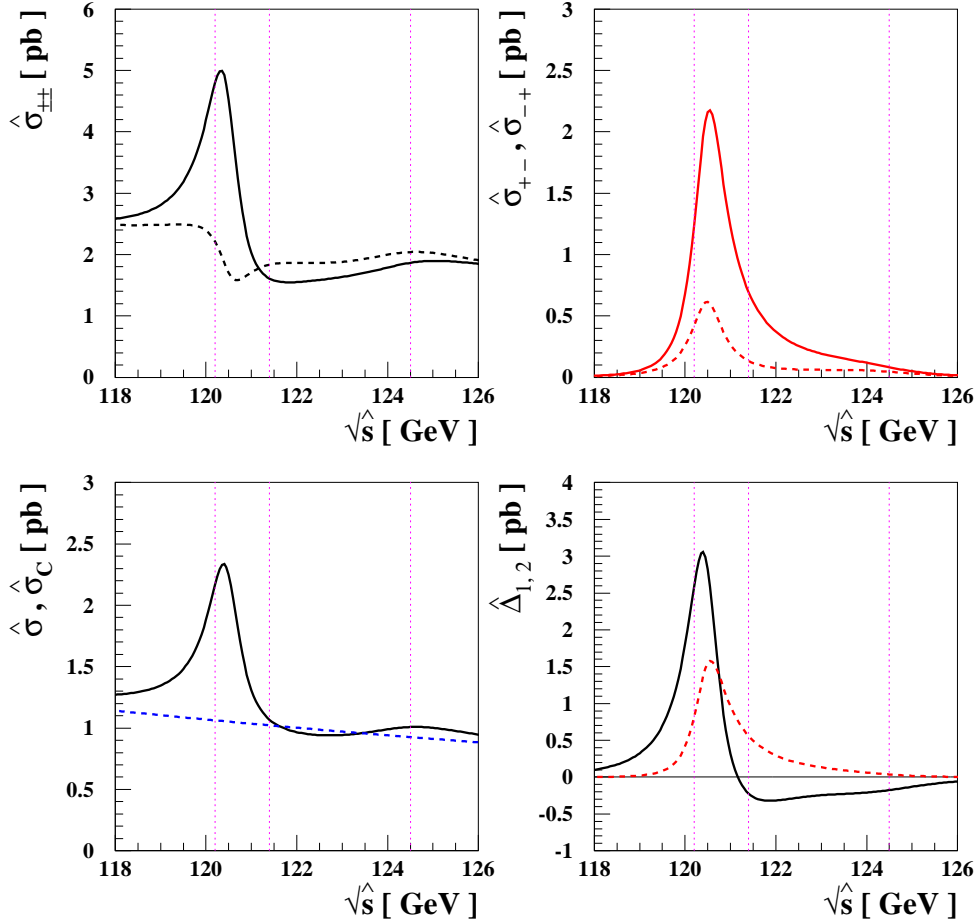,height=14cm,width=14cm}}
\vspace{-0.9 cm}
\caption{{\it 
The cross sections for the process $\gamma (\lambda)\gamma (\lambda)
\rightarrow \tau^+ (\sigma)\tau^- (\sigma)$ (Case I) in the three-way
mixing scenario with $\Phi_3=-10^\circ$ and $\Phi_{A_{t,b}}=90^\circ$, as
functions of $\sqrt{\hat{s}}$. In panels (a, b, c) and (d), the solid
lines are for $\hat{\sigma}_{++}$, $\hat{\sigma}_{+-}$, $\hat{\sigma}$ and
$\hat\Delta_1$, and the dashed lines for $\hat{\sigma}_{--}$,
$\hat{\sigma}_{-+}$, the continuum cross section $\hat{\sigma}_C$ and
$\hat\Delta_2$, respectively.  The three Higgs masses are indicated by 
vertical lines. 
}}
\label{fig:c1.3way}
\end{figure}

In Fig.~\ref{fig:c1.3way}, we show the cross sections for Case I (identical
photon and fermion helicities) as functions of the invariant mass of 
the tau leptons in the three-way mixing scenario. The upper two
frames are for $\hat{\sigma}_{\sigma\lambda}$.
When $\sigma=\lambda$ (upper-left frame), the cross sections are around 2-5 pb
and the contribution from the QED continuum is comparable to that
from the Higgs-mediated process around $\sqrt{\hat{s}}=120$ GeV, as seen from
$\hat{\sigma}_C$ (dashed line) in the lower-left frame. 
We observe the sizable difference
between $\hat{\sigma}_{++}$ and $\hat{\sigma}_{--}$, which is just the
$\hat\Delta_1$ shown as a solid line in the lower-right frame.
When $\sigma\neq\lambda$ (upper-right frame), the cross sections are 
peaked between $M_{H_1}$ and $M_{H_2}$  
with sizes of about 2 pb ($\hat{\sigma}_{+-}$) and 
0.5 pb ($\hat{\sigma}_{-+}$) as shown by solid and dashed lines,
respectively. In this case the contribution from the QED continuum
is negligible since the amplitude is suppressed by the factor
$(1-\beta_\tau)$, see Eq.~(\ref{eq:slC}). Again we see a sizable 
difference
between the two cross sections, which is the CP-violating cross section
$\hat\Delta_2$ shown in the frame below as a dashed line.

\begin{figure}[htb]
\vspace{-1.5cm}
\centerline{\epsfig{figure=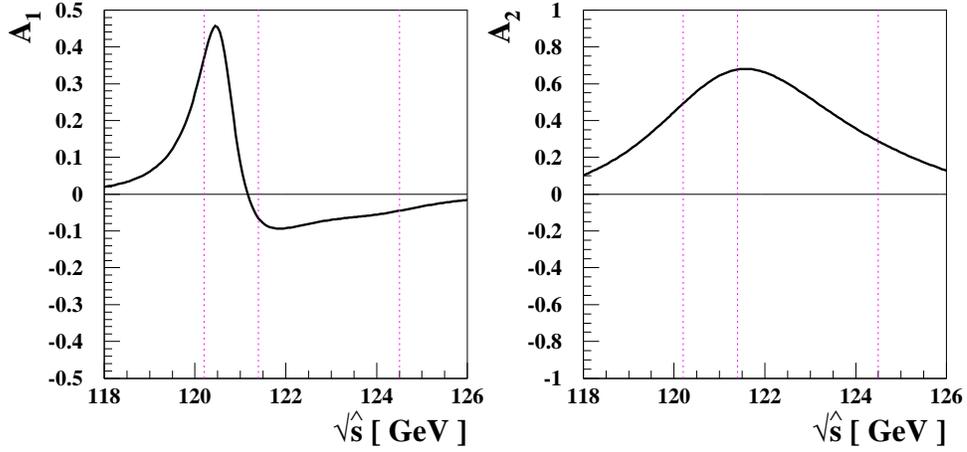,height=14cm,width=14cm}}
\vspace{-6.9 cm}
\caption{{\it 
The CP asymmetries ${\cal A}_1$ and ${\cal A}_2$ for the process 
$\gamma \gamma \rightarrow \tau^+ \tau^- $ in the three-way
mixing scenario with $\Phi_3=-10^\circ$ and $\Phi_{A_{t,b}}=90^\circ$, as
functions of $\sqrt{\hat{s}}$. 
}}
\label{fig:asym12}
\end{figure}

We show in Fig.~\ref{fig:asym12} the CP asymmetries defined by
\begin{equation}
{\cal A}_1\equiv
\frac{\hat\Delta_1}{\hat{\sigma}_{++}+\hat{\sigma}_{--}}\,, \ \ \ \
{\cal A}_2\equiv
\frac{\hat\Delta_2}{\hat{\sigma}_{+-}+\hat{\sigma}_{-+}}\,.
\label{eq:cpasym}
\end{equation}
The CP asymmetry ${\cal A}_1$ is larger than 10 \% in the region
$\sqrt{\hat{s}} \sim M_{H_1}$.
Assuming 10,000 $\gamma\gamma\to\tau^+\tau^-$ events after some experimental 
cuts to reconstruct the tau leptons and their polarizations, CP 
asymmetries larger than 1 \% could be measured.
The intrinsic CP asymmetry ${\cal A}_2$ is larger than 10 \% in the whole 
region and can
be as large as 70 \%.  Between $\sqrt{\hat{s}}=M_{H_1}$ and 
$M_{H_2}$ where
the sum of the cross sections $\hat{\sigma}_{+-}+\hat{\sigma}_{-+}$ is 
larger than 1 pb, the asymmetry is larger than 50 \%. However, the 
$\tau^+ \tau^-$ mass resolution is expected, realistically, to be larger 
than that for ${\bar b} b$ final states, so that these fine details would 
be washed out and only asymmetries integrated over the resonance peaks 
would be observable.


\begin{figure}[htb]
\vspace{-1.5cm}
\centerline{\epsfig{figure=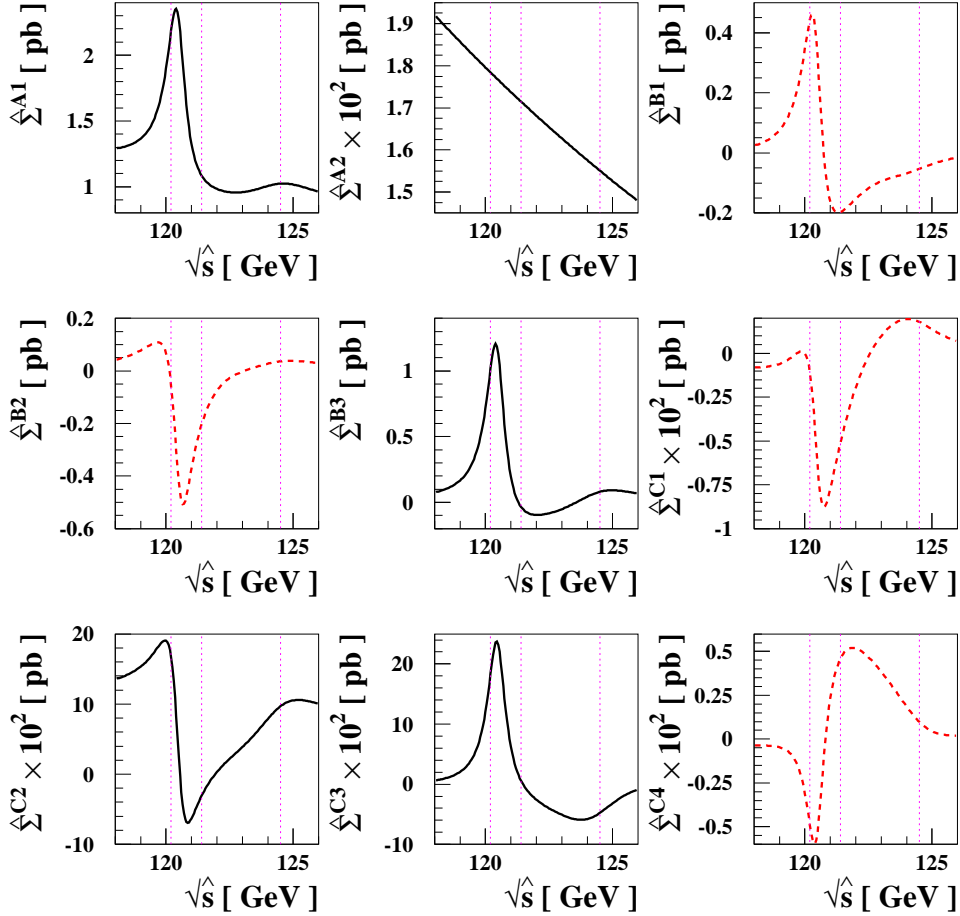,height=14cm,width=14cm}}
\vspace{-0.9 cm}
\caption{{\it The cross sections $\hat\Sigma^X$ for the process 
$\gamma (\lambda_1)\gamma (\lambda_2) 
\rightarrow \tau^+ (\sigma)\tau^- (\sigma)$ 
(Case II) in the three-way mixing 
scenario with $\Phi_3=-10^\circ$ and $\Phi_{A_{t,b}}=90^\circ$ as functions of
$\sqrt{\hat{s}}$. 
 }}
\label{fig:sig.abc}
\end{figure}
\begin{figure}[htb]
\vspace{-1.5cm}
\centerline{\epsfig{figure=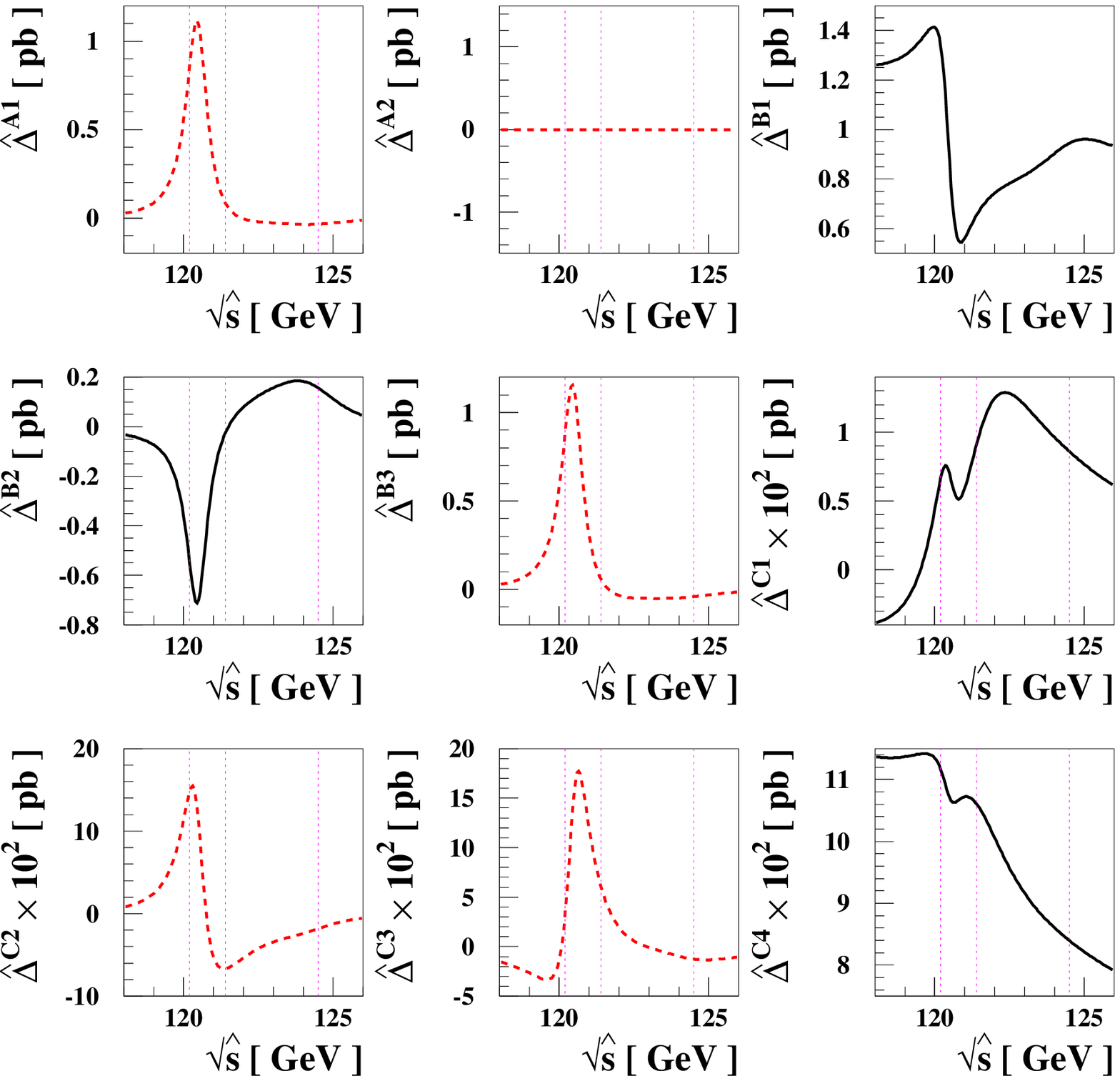,height=14cm,width=14cm}}
\vspace{-0.9 cm}
\caption{{\it The cross sections $\hat\Delta^X$ for the process 
$\gamma (\lambda_1)\gamma (\lambda_2) 
\rightarrow \tau^+ (\sigma)\tau^- (\sigma)$ 
(Case II) in the three-way mixing 
scenario with $\Phi_3=-10^\circ$ and $\Phi_{A_{t,b}}=90^\circ$ as functions of
$\sqrt{\hat{s}}$. 
 }}
\label{fig:del.abc}
\end{figure}

Figs.~\ref{fig:sig.abc} and ~\ref{fig:del.abc} display the cross sections
$\hat\Sigma^{A_i,B_j,C_k}(\gamma\gamma\rightarrow\tau^+\tau^-)$ and
$\hat\Delta^{A_i,B_j,C_k}(\gamma\gamma\rightarrow\tau^+\tau^-)$, respectively,
in the three-way mixing scenario. The CP-odd cross sections are shown with
dashed lines. 
Note that $\hat\Sigma^{A_2}$ and all the observables related with $C_i$ 
($\hat\Sigma^{C_k}$ and  $\hat\Delta^{C_k}$ with $k=1,2,3,4$) 
are one or two orders of magnitude 
smaller than other observables.
We see that $\hat\Sigma^{C_1}$ and $\hat\Sigma^{C_4}$, which are
CP-odd, are too small to be observed. But the other CP-odd observables
$\hat\Delta^{A_1}$, $\hat\Sigma^{B_1}$,
$\hat\Sigma^{B_2}$ and
$\hat\Delta^{B_3}$, which are larger than $0.1$ pb, may well be 
measurable.


\begin{figure}[htb]
\vspace{-1.5cm}
\centerline{\epsfig{figure=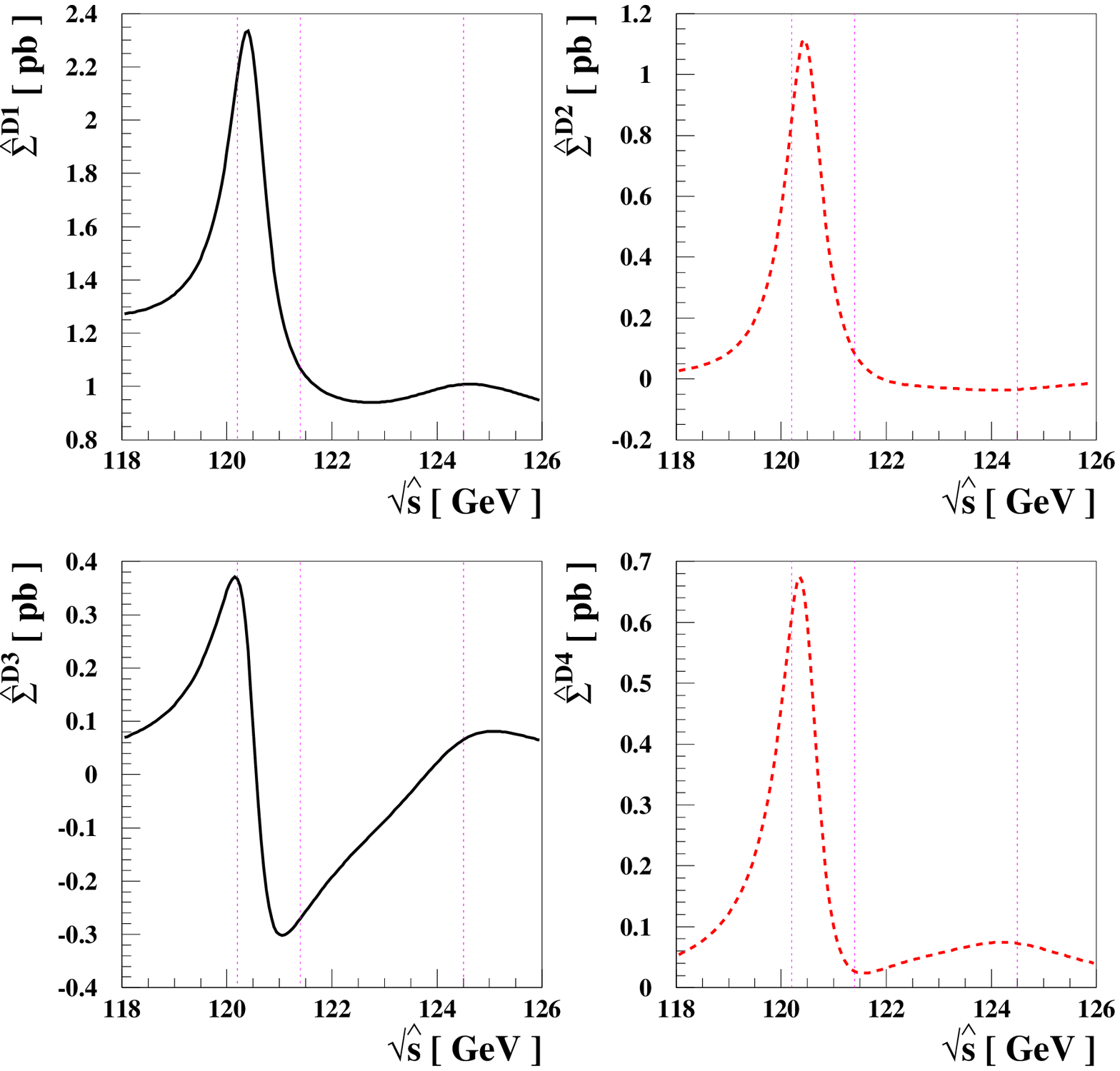,height=14cm,width=14cm}}
\vspace{-0.9 cm}
\caption{{\it The cross sections $\hat\Sigma^{D_i}$ for the process 
$\gamma (\lambda)\gamma (\lambda) 
\rightarrow \tau^+ (\bar{\sigma})\tau^- (\sigma)$ 
(Case III) in the three-way mixing 
scenario with $\Phi_3=-10^\circ$ and $\Phi_{A_{t,b}}=90^\circ$ as functions of
$\sqrt{\hat{s}}$. 
 }}
\label{fig:sig.d}
\end{figure}

\begin{figure}[htb]
\vspace{-1.5cm}
\centerline{\epsfig{figure=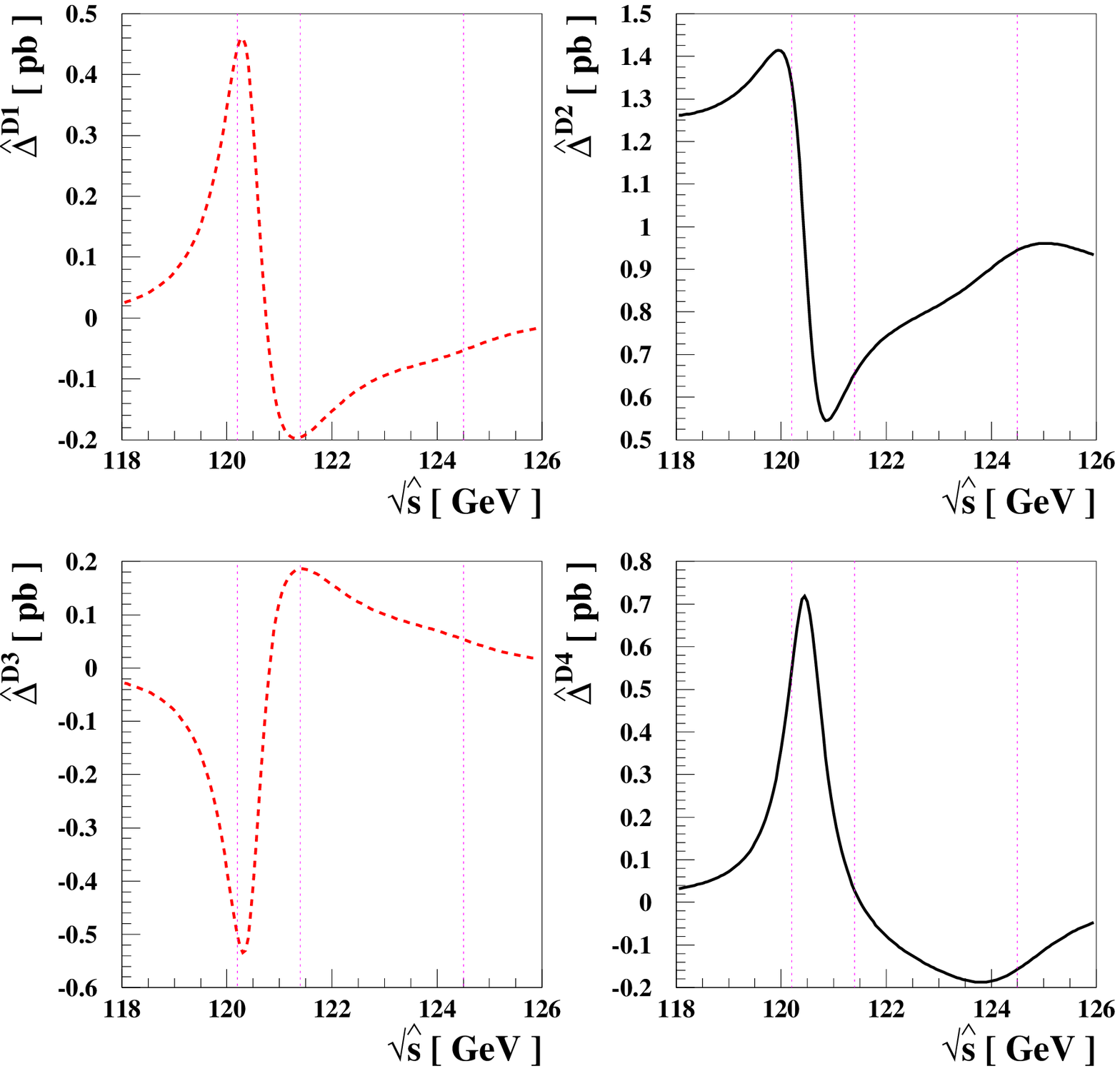,height=14cm,width=14cm}}
\vspace{-0.9 cm}
\caption{{\it The cross sections $\hat\Delta^{D_i}$ for the process 
$\gamma (\lambda)\gamma (\lambda) 
\rightarrow \tau^+ (\bar{\sigma})\tau^- (\sigma)$ 
(Case III) in the three-way mixing 
scenario with $\Phi_3=-10^\circ$ and $\Phi_{A_{t,b}}=90^\circ$ as functions of
$\sqrt{\hat{s}}$. 
 }}
\label{fig:del.d}
\end{figure}

Figs.~\ref{fig:sig.d} and ~\ref{fig:del.d} display the cross sections
$\hat\Sigma^{D_i}(\gamma\gamma\rightarrow\tau^+\tau^-)$ and
$\hat\Delta^{D_i}(\gamma\gamma\rightarrow\tau^+\tau^-)$, respectively, in the
three-way mixing scenario. Again, the CP-odd cross sections are indicated
by dashed lines. The CP-odd cross sections larger than $0.1$ pb
may well be measurable.

\setcounter{equation}{0}
\section{CP Violation in $\gamma \gamma \to {\bar t} t$ in a Two-Way 
Mixing\newline Scenario}
\label{sec:twocalx}

The formalism developed above applies equally well to the process $\gamma
\gamma \to {\bar t} t~$\cite{ERI,PPTT,CKLZ}. 
However, resonant ${\bar t} t$ production is not
possible in the scenario with strong three-way mixing that was presented
previously. Hence, in order to exhibit the possible CP-violating
signatures in $\gamma \gamma \to {\bar t} t$, we introduce a scenario with
a heavier charged Higgs boson and strong two-way neutral-Higgs mixing.

The parameters are taken as:
\begin{eqnarray}
\tan\beta&=&10\,, \ \ M_{H^\pm}^{\rm pole}=0.5\,{\rm TeV}\,, \ \
|\mu|=1\,{\rm TeV}\,, \nonumber \\
|M_3|&=&1\,{\rm TeV}\,, \ \ M_{\rm SUSY}=0.5\,{\rm TeV}\,, \ \
|A_{t,b}|=1\,{\rm TeV}\,, \nonumber \\
\Phi_{A_{t,b}}&=&90^\circ\,, \ \ \Phi_3=180^\circ\,.
\label{eq:toptop}
\end{eqnarray}
In this scenario, the masses and widths of the heavier neutral Higgs 
bosons are
\begin{eqnarray}
M_{H_2}&=&490.7\,{\rm GeV}\,, \ \ \Gamma_{H_2}=2.036\,{\rm GeV}\,,
\nonumber \\
M_{H_3}&=&495.2\,{\rm GeV}\,, \ \ \Gamma_{H_2}=1.969\,{\rm GeV}\,,
\end{eqnarray}
with $M_{H_1}=121.0$ GeV. 

\begin{figure}[htb]
\vspace{-1.5cm}
\centerline{\epsfig{figure=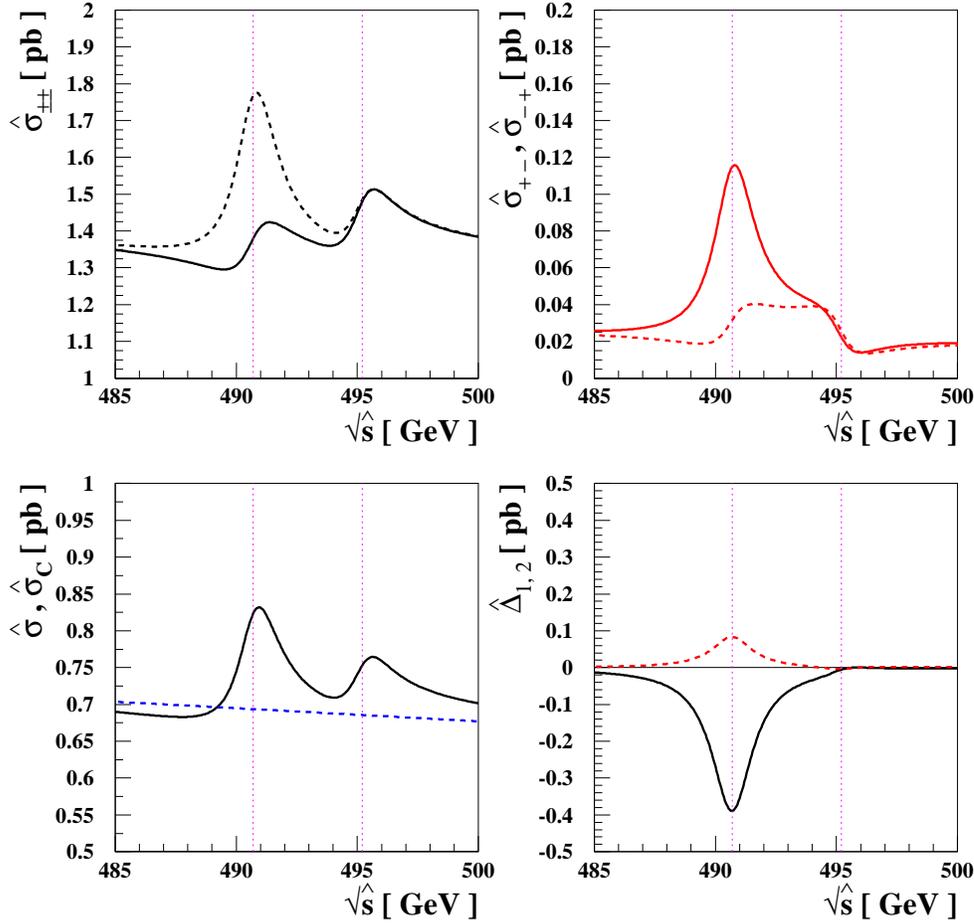,height=14cm,width=14cm}}
\vspace{-0.9 cm}
\caption{{\it The cross sections for the process 
$\gamma (\lambda)\gamma (\lambda) \rightarrow t (\sigma)\bar{t} (\sigma)$ 
(Case I) as functions of $\sqrt{\hat{s}}$. 
The parameter set (\ref{eq:toptop}) is taken.
The solid lines are for
$\hat{\sigma}_{++}$, $\hat{\sigma}_{+-}$, $\hat{\sigma}$, and $\hat\Delta_1$, and
the dashed lines for $\hat{\sigma}_{--}$, $\hat{\sigma}_{-+}$, 
the continuum cross section $\hat{\sigma}_C$, and $\hat\Delta_2$. 
The heavier Higgs-boson masses are represented with 
vertical lines.  
}}
\label{fig:c1.toptop}
\end{figure}

\begin{figure}[htb]
\vspace{-1.5cm}
\centerline{\epsfig{figure=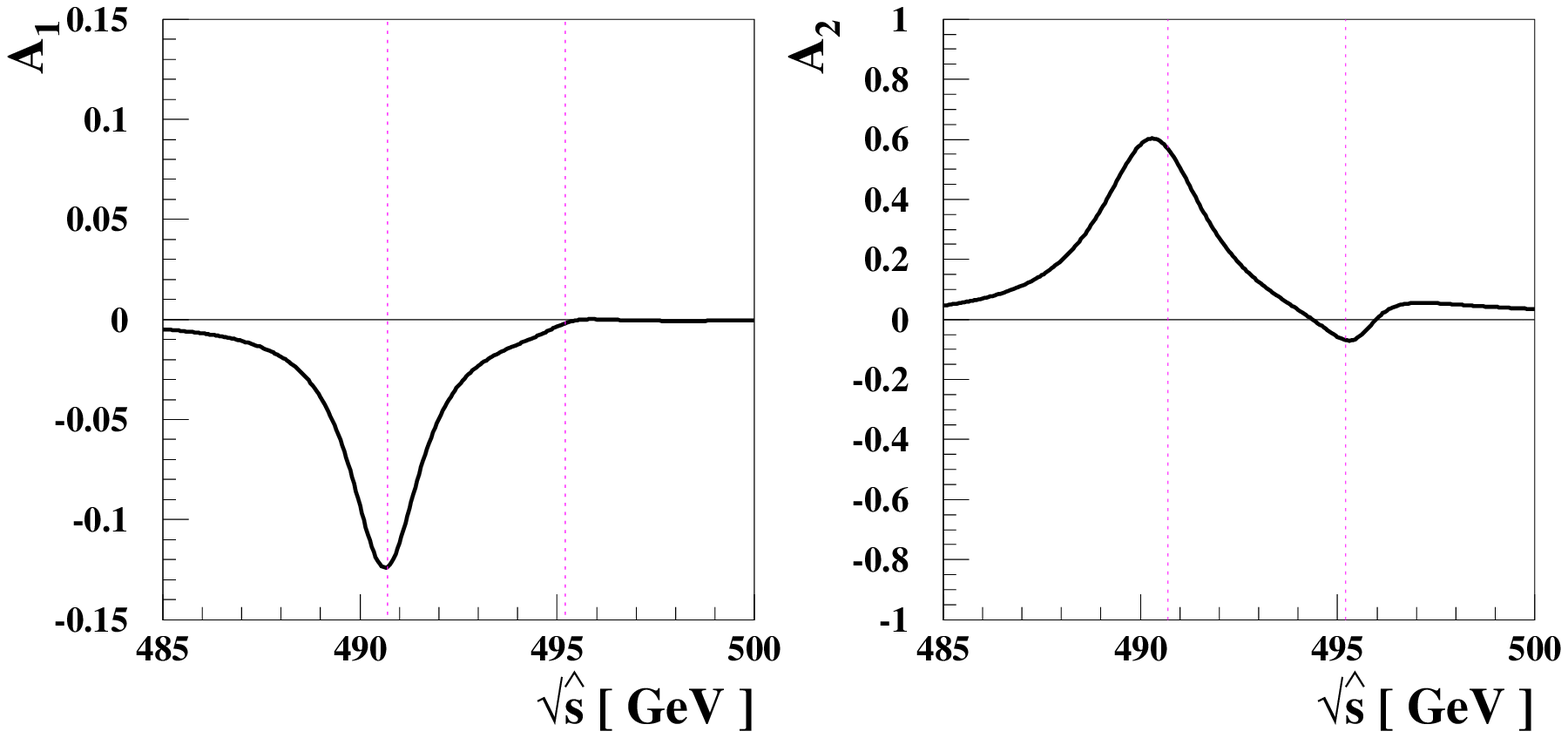,height=14cm,width=14cm}}
\vspace{-6.9 cm}
\caption{{\it 
The CP asymmetries ${\cal A}_1$ and ${\cal A}_2$ for the process 
$\gamma \gamma \rightarrow t \bar{t} $ in the two-way
mixing scenario as functions of $\sqrt{\hat{s}}$. 
}}
\label{fig:asym12.toptop}
\end{figure}

In Fig.~\ref{fig:c1.toptop}, we show the cross sections
$\hat{\sigma}_{\sigma\lambda}$, the helicity-averaged cross section
$\hat{\sigma}$, the QED continuum cross section $\hat{\sigma}_C$, and
CP-violating cross section $\hat\Delta_{1,2}$ as functions of the invariant mass
of a top quark pair, $\sqrt{\hat{s}}$. The line conventions are the same as in
Fig.~\ref{fig:c1.3way}. For $\sigma=\lambda$, the cross sections lie between
1.3 pb and 1.8 pb. The difference between two cross sections, $\hat\Delta_1$,
is about 0.4 pb at ${\sqrt{\hat s}}=M_{H_2}$, but is much smaller 
at ${\sqrt{\hat s}}=M_{H_3}$, as shown in the lower-right frame. 
When $\sigma\neq\lambda$, the cross sections are about an order of magnitude
smaller than those for the case $\sigma=\lambda$. The difference of the cross
sections, $\hat\Delta_2$, could be as large as $0.1$ pb at 
${\sqrt{\hat s}}=M_{H_2}$. The cross sections may well be
large enough to be measurable.
Two CP asymmetries ${\cal A}_{1,2}$ which are defined in 
(\ref{eq:cpasym}), are shown in Fig.~\ref{fig:asym12.toptop}. 
We see large CP asymmetries only around the $H_2$ Higgs-boson peak for the
parameter set chosen.


\begin{figure}[htb]
\vspace{-1.5cm}
\centerline{\epsfig{figure=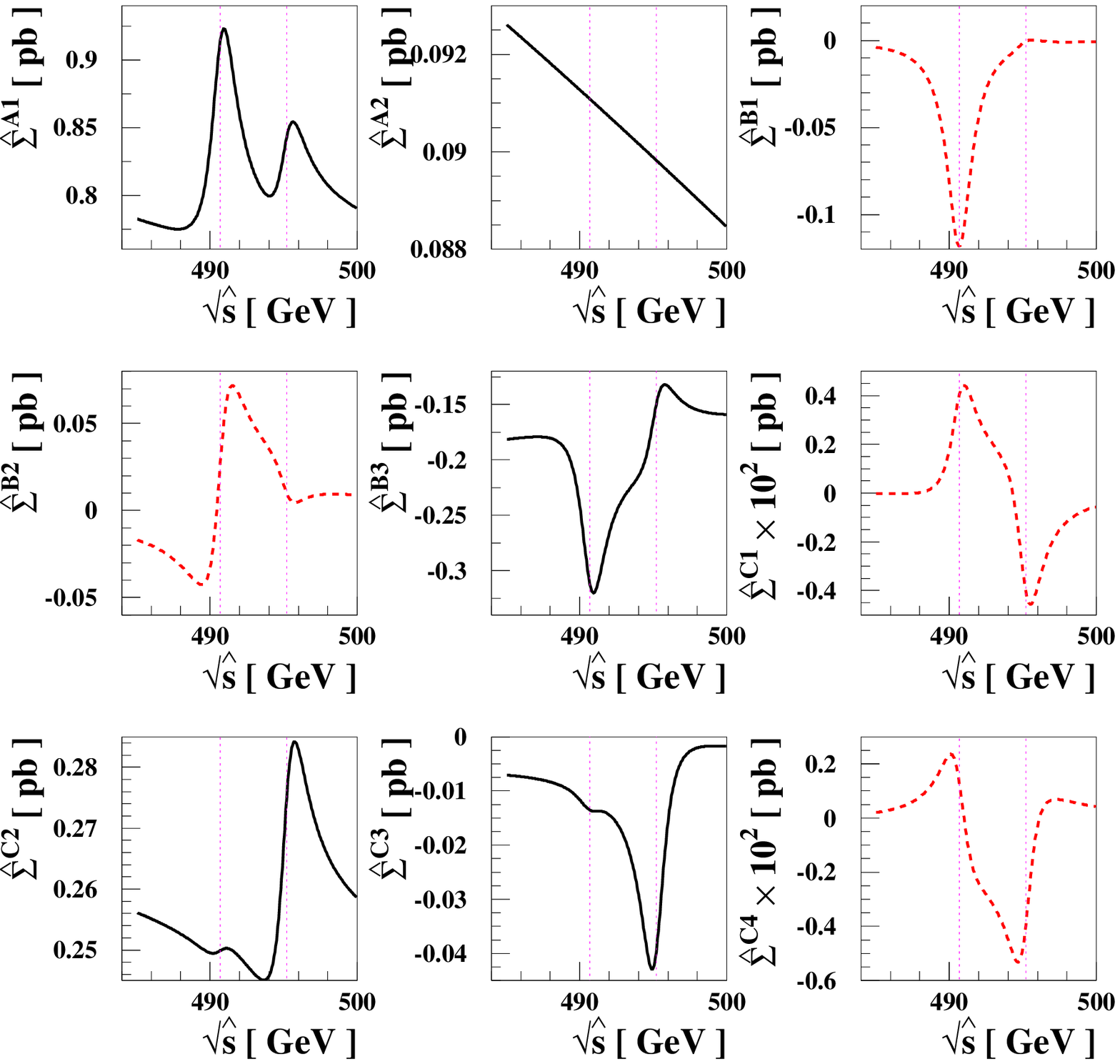,height=14cm,width=14cm}}
\vspace{-0.9 cm}
\caption{{\it The cross sections $\hat\Sigma^X
(\gamma (\lambda_1)\gamma (\lambda_2) 
\rightarrow t (\sigma)\bar{t} (\sigma))$ 
(Case II) as functions of $\sqrt{\hat{s}}$ 
for the parameter set (\ref{eq:toptop}).
 }}
\label{fig:sig.abc.toptop}
\end{figure}

\begin{figure}[htb]
\vspace{-1.5cm}
\centerline{\epsfig{figure=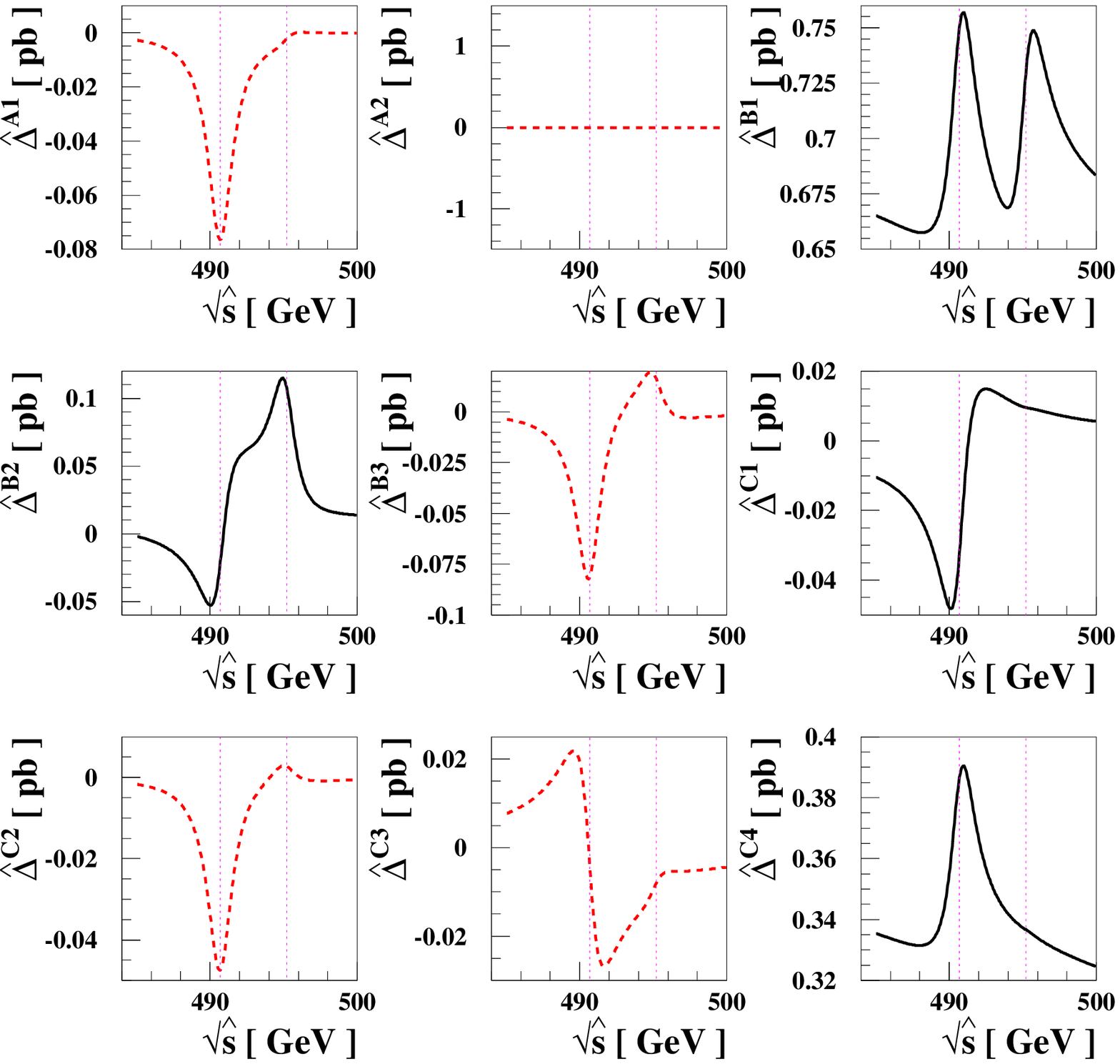,height=14cm,width=14cm}}
\vspace{-0.9 cm}
\caption{{\it The cross sections $\hat\Delta^X
(\gamma (\lambda_1)\gamma (\lambda_2) 
\rightarrow t (\sigma)\bar{t} (\sigma))$ 
(Case II) as functions of $\sqrt{\hat{s}}$ 
for the parameter set (\ref{eq:toptop}).
 }}
\label{fig:del.abc.toptop}
\end{figure}

In Figs.~\ref{fig:sig.abc.toptop} and \ref{fig:del.abc.toptop}, the cross
sections
$\hat\Sigma^{A_i,B_j,C_k}(\gamma\gamma\to t {\bar t})$ and
$\hat\Delta^{A_i,B_j,C_k}(\gamma\gamma\to t {\bar t})$ are shown, respectively.
All the cross sections larger than 0.01 pb may well be measurable.
Even the smallest CP-odd cross sections $\hat\Sigma^{C_1}$ and $\hat\Sigma^{C_4}$,
which are denoted by dashed lines,
have a size of a few fb at $\sqrt{\hat s}=M_{H_2}$ and $M_{H_3}$.


\begin{figure}[htb]
\vspace{-1.5cm}
\centerline{\epsfig{figure=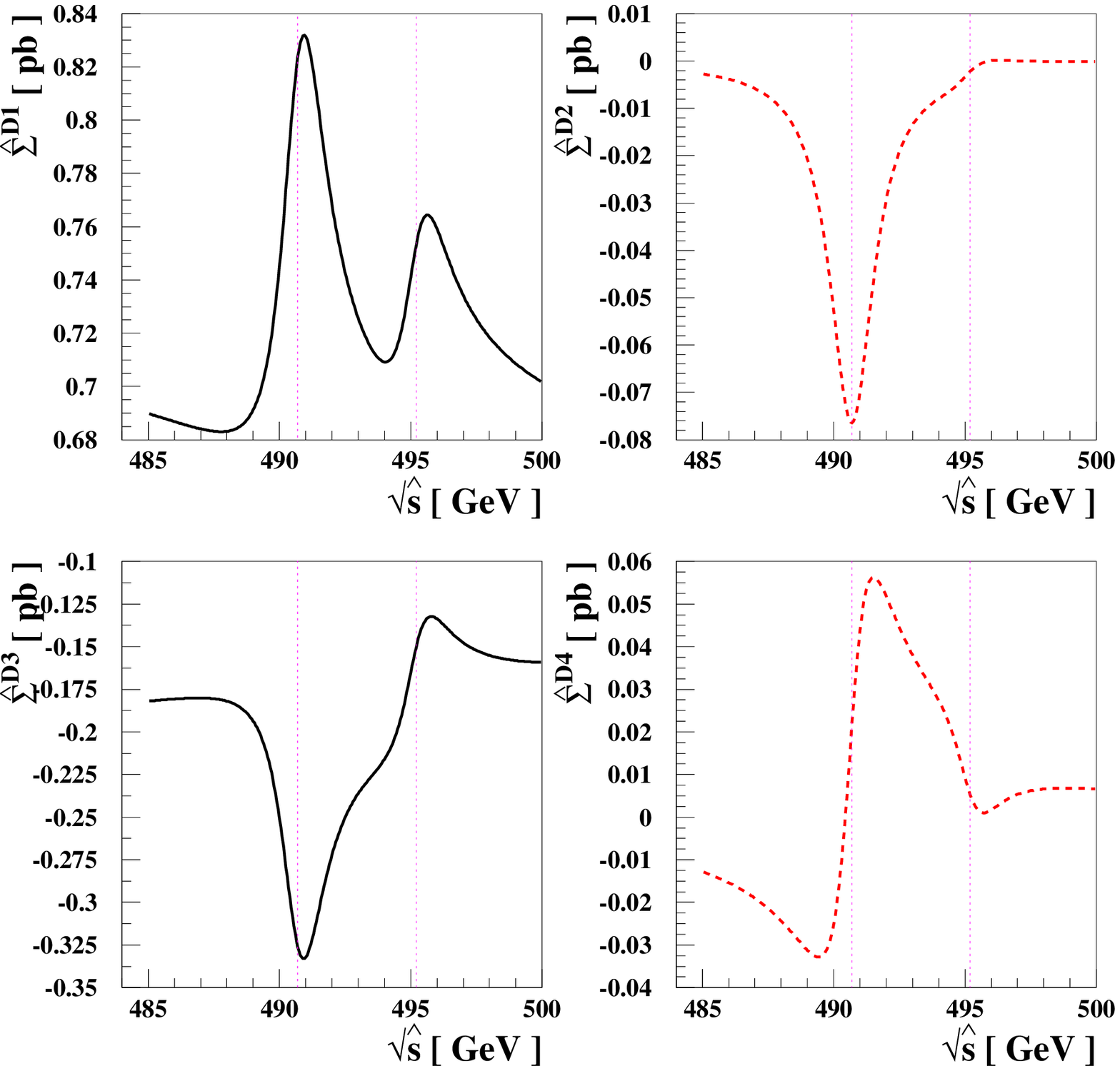,height=14cm,width=14cm}}
\vspace{-0.9 cm}
\caption{{\it The cross sections $\hat\Sigma^{D_i}
(\gamma (\lambda)\gamma (\lambda)
\rightarrow t (\sigma)\bar{t} (\bar{\sigma}))$
(Case III) for the parameter set (\ref{eq:toptop}) as
functions of $\sqrt{\hat{s}}$.
 }}
\label{fig:sig.d.toptop}
\end{figure}

\begin{figure}[htb]
\vspace{-1.5cm}
\centerline{\epsfig{figure=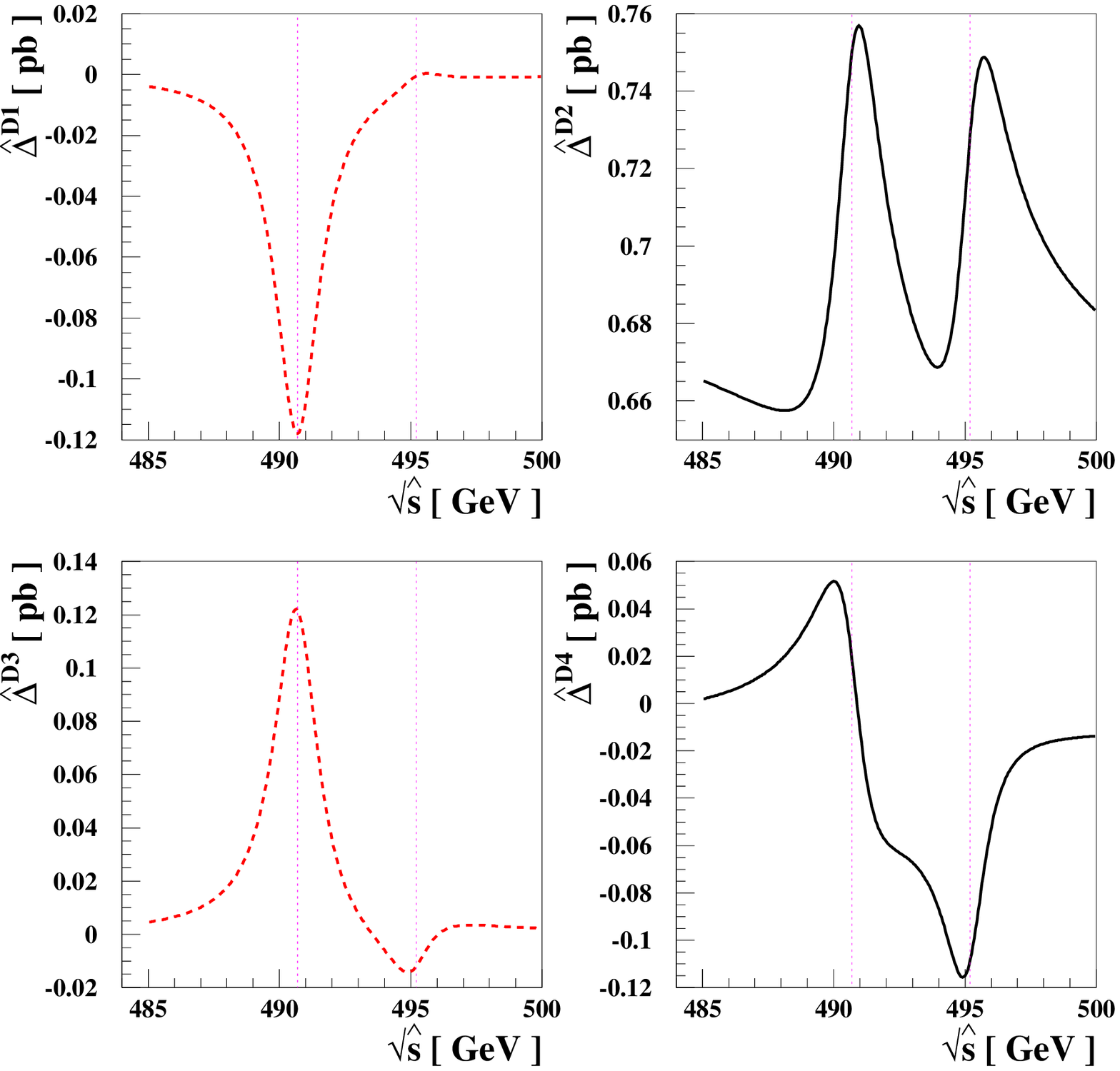,height=14cm,width=14cm}}
\vspace{-0.9 cm}
\caption{{\it The cross sections $\hat\Delta^{D_i}
(\gamma (\lambda)\gamma (\lambda)
\rightarrow t (\sigma)\bar{t} (\bar{\sigma}))$
(Case III) for the parameter set (\ref{eq:toptop}) as
functions of $\sqrt{\hat{s}}$.
 }}
\label{fig:del.d.toptop}
\end{figure}

Figs.~\ref{fig:sig.d.toptop}, and \ref{fig:del.d.toptop} display the
cross sections $\hat\Sigma^{D_i}(\gamma\gamma\rightarrow t{\bar t})$ and
$\hat\Delta^{D_i}(\gamma\gamma\rightarrow t{\bar t})$, respectively, in the
two-way mixing scenario. Again, the CP-odd cross sections are indicated
by dashed lines, and the CP-odd cross sections are large enough to
be measured.


\setcounter{equation}{0}
\section{Conclusions}
\label{sec:conx}

There is general agreement that the CP violation contained in the Standard 
Model, though it is consistent with all the laboratory data available so 
far, is inadequate for generating the baryon asymmetry of the Universe. 
One of the most appealing scenarios for physics beyond the Standard Model 
that might yield sufficient supplementary CP violation is supersymmetry 
which reopens the possibility of baryogenesis at the electroweak
scale~\cite{EWBAU}. 
Supersymmetric CP violation may appear directly both in sparticle 
production or decays and in the production and decays of MSSM Higgs 
bosons.

We presented previously a general  formalism for analyzing CP-violating
phenomena in the  production, mixing and decay  of a coupled system of
multiple CP-violating  neutral Higgs bosons, and applied it to Higgs 
production and decay at the LHC~\cite{ELP1}.

In this paper, as a further application of this formalism, we have studied
in detail the production and decays of CP-violating MSSM $H_{1,2,3}$
bosons in $\gamma \gamma$ collisions, studying $\mu^+\mu^-$,
$\tau^+ \tau^-$, ${\bar b} b$ and ${\bar t} t$ final states. 
We have constructed more than 20 independent observables by exploiting the
controllable beam polarization at $\gamma\gamma$ colliders and the possibly
measurable final-fermion polarizations. We have classified them
according to their CP and CP$\tilde{\rm T}$ parities and note that the half of
them are genuine CP-odd observables.
We have considered two specific MSSM
scenarios that predict either (a) three nearly degenerate, strongly-mixed
Higgs bosons with $M_{H_{1,2,3}} \sim 120$~GeV or (b) two nearly
degenerate, strongly-mixed Higgs bosons with $M_{H_{2,3}} \sim 490$~GeV.
Some of the CP-violating signatures we have explored may be quite large,
rising to 20~\% in some cases. Particularly promising examples seem to be
the asymmetry ${\cal A}_{0}$ in $\gamma \gamma \rightarrow {\bar b}b$ and
$\gamma \gamma \rightarrow \mu^+\mu^-$ in
the three-way mixing scenario, and the asymmetries ${\cal A}_{1}$ and
${\cal A}_{2}$ for the processes $\gamma \gamma \rightarrow \tau^+ \tau^-$
in the three-way mixing scenario and $\gamma \gamma \rightarrow {\bar t}t$
in the two-way mixing scenario. However, many other potential signatures
may also be interesting.

Our study confirms that a $\gamma \gamma$ collider would be a valuable
tool for unravelling CP violation in the MSSM. In particular, the
controllable initial-state polarizations offered by a $\gamma \gamma$
collider could provide sensitive tools for unravelling the origin of CP
violation and, by extension, the possibility of electroweak baryogenesis.

\subsection*{Acknowledgements}
We thank Klaus Desch for valuable advice on experimental parameters relevant
for this study.
The work of JSL and AP is supported in part by the PPARC research
grant PPA/G/O/ 2000/00461.

\newpage


\clearpage
\noindent

\end{document}